\documentclass[12pt,preprint]{aastex}
\input colordvi
\newcommand{\ltwid}{\mathrel{\raise.3ex\hbox{$<$\kern-.75em\lower1ex\hbox{$\sim$
}}}}
\newcommand{\gtwid}{\mathrel{\raise.3ex\hbox{$>$\kern-.75em\lower1ex\hbox{$\sim$
}}}}
\newcommand{\pr}{{\s\prime}}
\newcommand{\be}{{\bf e}}
\newcommand{\s}{\scriptscriptstyle}
\begin{document}
\begin{center}{\bf The Free Precession and Libration of Mercury}\\
\vspace{.5in}
S. J. Peale\\
Department of Physics\\
University of California\\
Santa Barbara, CA 93106\\
email: peale@io.physics.ucsb.edu\end{center}

%
%
\begin{center}{\bf Abstract}\end{center}
An analysis based on the direct torque equations including tidal
dissipation and a viscous core-mantle coupling is used to determine 
the damping time scales of O($10^5$) years for free precession of the
spin about the Cassini state and free libration in longitude for
Mercury.  The core-mantle coupling dominates the damping over the
tides by one to two orders of magnitude for the plausible parameters
chosen. The short damping times compared with the age of the solar
system means we must find recent or on-going excitation mechanisms if
such free motions are found by the current radar experiments or the
future measurement by the MESSENGER and BepiColombo spacecraft that
will orbit Mercury. We also show that the average precession rate is
increased by about 30\% over that obtained from the traditional
precession constant because of a spin-orbit resonance induced
contribution by the $C_{22}$ term in the expansion of the
gravitational field.  The $C_{22}$ contribution also causes the path
of the spin during the precession to be slightly elliptical with a
variation in the precession rate that is a maximum when the obliquity is
a minimum.  An observable free precession will compromise the
determination of obliquity of the Cassini state and hence of $C/M_{\s M}R^2$
for Mercury, but a detected free libration will not compromise the
determination of the forced libration amplitude and thus the
verification of a liquid core.



\section{Introduction \label{sec:intro}}
A major goal of the MESSENGER (Solomon, et al. 2001) and
BepiColombo (Anselmi \& Scoon, 2001) missions to Mercury and
ground-based Radar Speckle Displacement Interferometry (RSDI)
observations (Holin, 1988, 1992, 2003; Margot {\it et al.}
2003) is the determination of the structure 
and state of Mercury's interior.  The probe of interior properties is
based on the accurate determination of the four parameters,
$i,\,\phi_0,\,J_2$ and $C_{22}$, where $i$ is Mercury's
obliquity, $\phi_0$ is the amplitude of the physical libration in
longitude, and $J_2$ and $C_{22}$ are the second degree harmonic
coefficients in the expansion of Mercury's gravitational field.  These
parameters are sufficient for determining the factors in the equation
(Peale, 1976a; 1981; 1988; Peale {\it et al.} 2002)  
\begin{equation}
\left(\frac{C_m}{B-A}\right)\left(\frac{B-A}{M_{\s M}R^2}\right)\left(
\frac{M_{\s M}R^2}{C}\right)=\frac{C_m}{C}\leq 1, \label{eq:cmc}
\end{equation}
where $M_{\s M}$ and $R$ are the mass and radius of Mercury, $A<B<C$ are the
principal moments of inertia, and $C_m$ is the polar moment of inertia
of the mantle alone. 

\begin{figure}[h]
\plotone{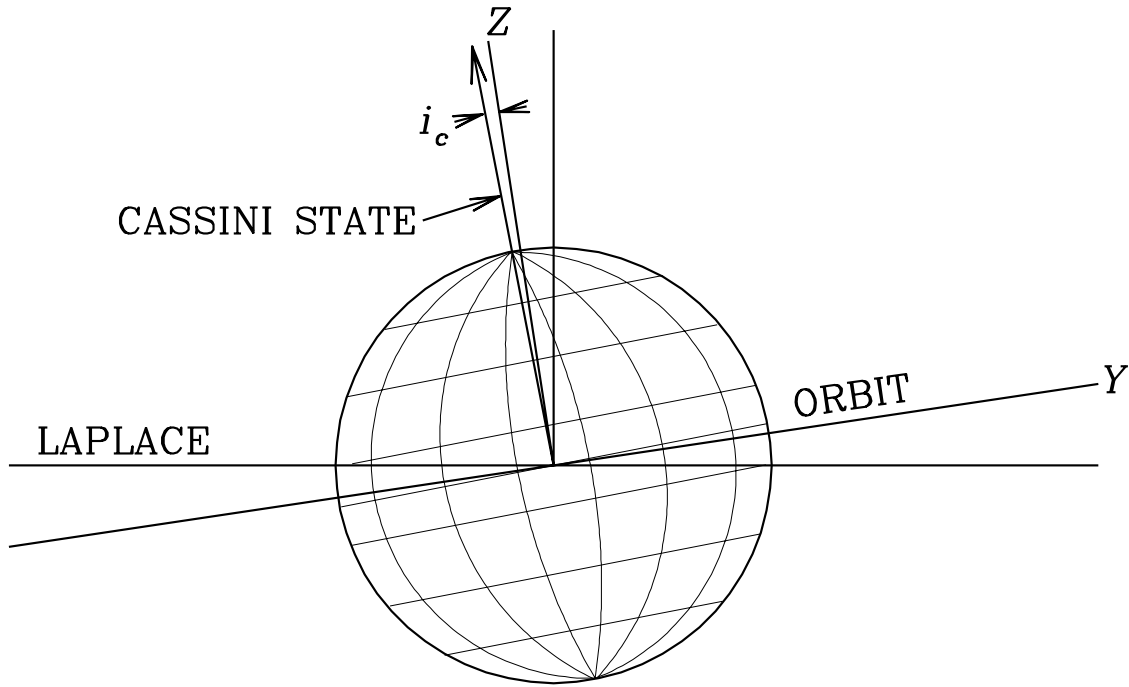}
\caption{Geometry of Cassini state 1. The $Z$ axis is perpendicular to
the orbit plane, and the ascending node of the orbit on the Laplace plane
and the ascending node of the equator on the orbit plane both coincide with
the $X$ axis, which comes out of the paper. The obliquity of the
Cassini state is $i_c$. \label{fig:cassinistate}}
\end{figure}

The first factor follows from the amplitude of
the physical libration
\begin{equation}
\phi_0=\frac{3}{2}\left(\frac{B-A}{C_m}\right)\left(1-11e^2
+\frac{959}{48}e^4+\cdots\right),\label{eq:libamp}
\end{equation}
where $e$ is the orbital eccentricity, and $C_m$ occurs in the
denominator because a liquid core is not expected to follow the 88 day
physical libration of the mantle. The second factor follows from
\begin{equation} 
\frac{B-A}{M_{\s M}R^2}=4C_{22}, \label{eq:c22}
\end{equation}
and the third from (Peale, 1969)
\begin{equation}
\frac{C}{M_{\s M}R^2}=\displaystyle\frac{\left[J_2/(1-e^2)^{3/2}
+2C_{22}\displaystyle\left(\frac{7}{2}e-\frac{123}{16}e^3\right)\right]}
{(\sin{I})/i_c-\cos{I}}\frac{n}{\mu}, \label{eq:cmr2}
\end{equation}
where $i_c$ is the obliquity of Cassini state 1,
({\it E.g.}, Colombo, 1966:, Peale, 1969), which Mercury is expected to
occupy ($i=i_c$), $I$ is the inclination of the orbit plane to
the Laplace plane on which Mercury's orbit precesses at the constant
rate $-\mu$, and $n$ is the orbital mean motion, and the equation
assumes $i_c\ll 
1$. In a Cassini state the spin axis is coplanar with the normal to
the Laplace plane and the orbit normal as both the spin and the
orbit normal precess around the Laplace plane normal with an
approximately 280,000 year period. Fig. \ref{fig:cassinistate} shows
the configuration for Cassini state 1, where the ascending node of
the equator plane on the orbit plane remains coincident with the
ascending node of the 
orbit plane on the Laplace plane. In Fig. \ref{fig:cassinistate},
the $Z$ axis is normal to the orbit plane, the $Y$ axis is in the
orbit plane, and the $X$ axis comes out of the paper. 
Eq. (\ref{eq:cmr2}) shows the dependence of $C/M_{\s M}R^2$ on
$J_2,\,C_{22}$ and $i_c$, and together, the last three equations
show the necessity of knowing $i_c,\,\phi_0,\,J_2$ and $C_{22}$ to
determine $C_m/C$. 

{\bf [Figure 1]}

Conditions for the success of the experiment are \newline
1) The liquid core must not follow the mantle during the 88 day forced
libration in longitude.\newline
2) The core must follow the mantle during the 280,000 year orbit
precession, where the spin axis is locked to this precession if it
occupies the Cassini state.\newline
3) $B-A$ must be due to the mantle alone.\newline
4) Adiabatic invariance must keep the spin close to the Cassini state
during the slow variation of the orbital elements. 

The first two conditions are satisfied for viscous core-mantle
coupling, topographical coupling due to core-mantle boundary (CMB)
irregularities, magnetic core-mantle coupling, and gravitational
coupling between an axially asymmetric solid inner core and the mantle  
(Peale, et al. 2002). The third condition requires that there not be
long wavelength irregularities in the CMB such that dense core
material contributes to $B-A$. Such irregularities could be induced by
mantle convection (D. Stevenson, private communication, 2003), but a
major contribution to $B-A$ would occur
only if the convective cells have circumferential scales comparable to
the planet radius $R$.  Otherwise, the plus and minus contributions due to
short wavelength topography would tend to average to a zero total
contribution to $B-A$. As the iron core must be near $0.75R$ in radius
(Siegfried and Solomon, 1974), the horizontal scale of any convective
cells is unlikely to greatly exceed the $0.25R$ thickness of the
mantle. It has also been argued that Mercury's mantle convection
ceased early in Mercury's history (Reese and Peterson, 2002), and the
CMB should thereby be nearly axially symmetric with little or no
contribution to $B-A$ from the core.

It is implicitly assumed in the above that all free motions associated
with Mercury's spin are damped to near zero amplitude, and that the
spin remains close to the current location of Cassini state 1 in spite
of slow variation of the orbital parameters that define the state.
The free motions include a free libration in longitude, a free
precession of the spin about  
Cassini state 1, and a free wobble.  The free libration in longitude
results when the axis of minimum moment of inertia does not point
toward the Sun when Mercury is at perihelion.  The gravitational 
torque on the permanent asymmetry of the planet, averaged around the
orbit, tends to restore the alignment of the axis of minimum moment of
inertia with the Sun at perihelion because of the 3:2 spin-orbit
resonance, so viewed at perihelion, the long
axis will tend to librate about the direction toward the Sun with a
period of about 10 years.  The free precession of the spin axis is
like the 26,000 year precession of the Earth's spin axis about the
ecliptic, except the precession is about the Cassini state instead of
the orbit normal, where the deviation of the Cassini state from the
orbit normal is less than 2 arcmin. The period of this precession is
$O(1000\,{\rm years})$. Free wobble or non principal axis
rotation occurs when the spin axis does not coincide with the axis of
maximum moment of inertia.  The spin axis then precesses around the
axis of maximum moment of inertia in the body frame of reference with
a period $O(500\, {\rm years})$. The wobble is analogous to the Chandler
wobble of the Earth.  The angular momentum is conserved during free
wobble, and the spin axis makes only small excursions relative to
inertial space.  The spin angular momentum is not conserved during the
free libration and free precession as the planet responds to the
external gravitational torque due to the Sun.  In the former the spin
axis direction is maintained as the magnitude of the spin varies
periodically, whereas in the latter the spin describes a cone about
the Cassini state, which is almost a cone relative to inertial space. 

The term ``free'' characterizes
these motions as having arbitrary phases and amplitudes, which
amplitudes, if significantly different from zero, could complicate the
interpretation of the spacecraft and radar data constraining the
properties of the interior. For example, a finite amplitude free
precession would displace Mercury's spin vector from the Cassini state,
and its arbitrary phase and amplitude would thereby increase the
uncertainty in the obliquity of the state used in determining $C/M_{\s
M}R^2$
in Eq. (\ref{eq:cmr2}). Tidal friction has been shown to carry the
spin to Cassini state 1 from almost any initial configuration on
a time  scale that is short compared to the age of the solar system
(Peale, 1974; Ward, 1975).  Free
libration in longitude (9.2 year period, for $(B-A)/C_m=3.5\times 10^{-4}$)
is similarly damped, but such a free
libration does not confuse the experiment, as the 88 day forced
libration is superposed on the free libration (Peale, {\it et al.}
2002), and its amplitude is readily discernible.  Any free wobble will
also damp due to dissipation by internal friction as the equator plane
changes relative to the body principal axis system in addition to tidal
friction and core-mantle dissipation. For the Moon, tidal dissipation
is about 4 times more effective than the dissipation caused by the
changing orientation of the equator  (Peale, 1976b).  Finally it was
asserted (Peale, 
1974) that the spin would remain close to the instantaneous position
of the Cassini state if it were ever close because of the adiabatic
invariance of an action integral that is equivalent to the  solid
angle swept out by the spin vector as it precesses around the Cassini
state. 

If, on the other hand,  free motions of measurable amplitude are
found, excitation mechanisms must be sought given the damping. 
Relatively small amplitudes of free motions can be detected by orbiting 
spacecraft (Zuber and Smith, 2000) and radar (Holin, 2003; Margot et al. 
2003). The damping time scales constrain the types of excitation
mechanisms that are likely to be responsible. For example, short time
scales make excitation by the impact of large objects less likely (See
Section \ref{sec:impact}). The existence
of a liquid core introduces damping by core-mantle interaction that
shortens the time scales for the damping of free motions deduced for
tidal dissipation. This paper is the first of three where we consider
the free motions and quantify the adiabatic invariance of the spin
axis separation from the Cassini state.  Here we derive and compare
damping time scales for the two processes of tidal friction and
core-mantle dissipation with viscous coupling for free libration and
free precession.  Quantifying the adiabatic invariance and determining the
damping of a free wobble require different approaches, and these will
be considered in the following papers.  The damping of the free
rotational motions of the Moon have been determined earlier using a
Hamiltonian formulation with dissipation accounted for by introducing
phase lags in the periodic terms describing the response of the Moon
to various forcing potentials (Peale, 1976b). This analysis is general
albeit somewhat complicated.  For the libration and precession
considered here we use a much simpler approach by fixing the orbit and
writing the equations of motion directly in terms of the 
external and internal torques on the mantle and core.  However, this
simplified approach will not work well for the wobble, so the general
Hamiltonian analysis will be used in a following paper to treat the
wobble damping, although the torque equations will be applied there
for the core-mantle coupling.  In Sections \ref{sec:precession} and
\ref{sec:libration}, we 
show that the damping times for the free precession and the free
libration in longitude are both near $10^5$ years with the core-mantle
interaction being between one and two orders of magnitude more
effective than tidal friction. In Section \ref{sec:impact} we show that
the time between impacts capable of generating observable amplitudes
of free libration or precession exceeds $10^9$ years, making such
excitation exceeding unlikely.  Summary and conclusions follow in
Section \ref{sec:summary}.  We begin with
the free precession, as the equations describing the damping of the
libration in longitude are scalar versions of those describing the
damping of the precession. 

\section{Damping of free precession \label{sec:precession}}

Generally, the spin axis precesses about Cassini state 1 which is
offset approximately 1.7 arcmin from the orbit normal.  However, for the
purposes of calculating the damping, we assume the orbit is fixed, so
that the precession will be about the orbit normal. (The Cassini state
for a fixed orbit can be thought to coincide with the orbit normal,
since the state approaches the orbit normal as the orbital precession
rate is decreased, or it can be thought not to exist at all.)
Accordingly, we adopt an $XYZ$ inertial system centered on
Mercury with the XY plane coincident with the orbit plane, the $X$
axis pointing from Mercury to the perihelion of the Sun's relative
orbit, and with unit vector $\be_o=\be_{\s Z}$ being the orbit normal.
During the precession about the orbit normal, the damping
of the precession angle about the orbit normal will be essentially the
same as the damping of the precession about the Cassini state without
involving the complications induced by the orbit precession.  (See
Peale, (1976b) for a complete analysis applied to the Moon.)

We assume that the core mantle interaction is simply proportional to
the difference in the vector spin angular velocities of the 
mantle and core as the simplest possible coupling. This coupling is
consistent with a viscous coupling between two rigid spheres.  Mercury is
rotating slowly enough such that pressure coupling of the core and
mantle from the elliptical shape of the core-mantle boundary is not
important, which leaves viscous coupling the most probable.  We also assume
principal axis rotation.  The equations describing the motions of the
core and mantle are then 
\begin{eqnarray}
C_m\ddot{\vec\psi}_m&=&{\vec N}+{\vec T}-k(\dot{\vec\psi}_m-
\dot{\vec\psi}_c)\nonumber\\
C_c\ddot{\vec\psi}_c&=&k(\dot{\vec\psi}_m-\dot{\vec\psi}_c),
\label{eq:vectoreqns}
\end{eqnarray}
where $\dot{\vec\psi}_{m,c}$ are the angular velocities of the mantle
and core respectively, $C_{m,c}$ are the moments of inertia about the
spin axis of the mantle and core, $k$ is the coupling constant for the
core-mantle torque, $\vec N$ is the torque on Mercury's permanent
figure due to the Sun, and ${\vec T}$ is the tidal torque.
We first describe the free precession of Mercury, which is affected by
the spin-orbit resonance $\langle\dot\psi_m\rangle=1.5n$, where
$\langle\dot\psi_m\rangle$ is the average value of the magnitude of
$\dot{\vec\psi_m}$, and $n$ is the mean motion. 

\subsection{Free Precession \label{sec:freeprec}}

\begin{figure}[h]
\plotone{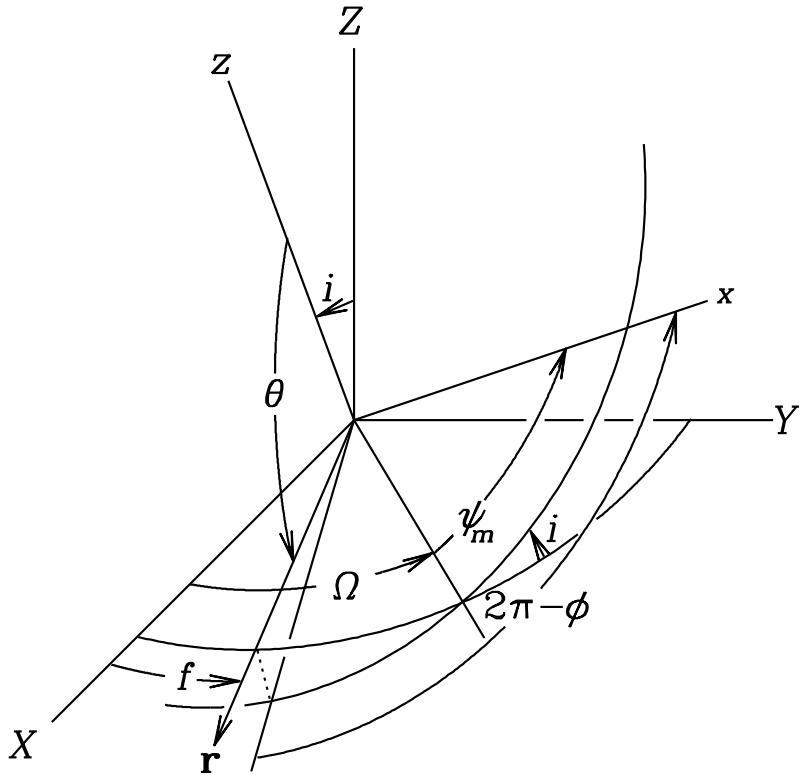}
\caption{Coordinates for describing Mercury's precession. The $XY$
plane is the orbit plane with the $X$ axis pointing toward the
perihelion position, $\Omega$ is the longitude of the ascending
node of the equator plane on the orbit plane, $i$ is the obliquity,
and $\psi_m$ locates the $x$ body axis of minimum moment of inertia
(of the mantle) relative to the ascending node. The $z$ axis is the
axis of maximum moment of inertia, which is also the spin
axis. Spherical polar coordinates $r\theta\phi$ locate the Sun
relative to the body axes.  \label{fig:orbitfixed}}  
\end{figure}

We start with the first of Eqs. (\ref{eq:vectoreqns}), with ${\vec
T}=k=0$ and keep only the second order terms in the expansion of
Mercury's gravitational potential $\Phi$. Then the torque on Mercury
is the negative of the torque on the Sun, so
\begin{eqnarray}
{\vec r}\times\nabla\Phi={\vec N}&=& -\frac{GM_{\s M}M_{\s\odot} R^2}{r^3}
\bigg[6C_{22}\sin{\theta}\sin{2\phi}\,{\bf e}_{\theta}\nonumber\\
&+&(3J_2\sin{\theta}\cos{\theta}+6C_{22}\sin{\theta}\cos{\theta}
\cos{2\phi})\,{\bf e}_{\phi}\bigg], \label{eq:torque}
\end{eqnarray} 
where $G$ is the gravitational constant, $M_{\s M}$ and $M_{\s\odot}$ are
the masses of Mercury and the Sun respectively, $R$ is Mercury's
radius, $J_2$ and $C_{\s 22}$ are second order 
harmonic coefficients in the expansion of Mercury's gravitational
field, $r\theta\phi$ are the spherical polar coordinates of the Sun
relative to the principal $xyz$ axis system with ${\bf e}_\theta$ and
${\bf e}_\phi$ being unit vectors in the $\theta$ and $\phi$
directions. The $z$ axis is the axis of maximum moment of inertia
coincident with the spin axis, and $x$ and $y$ are axes of minimum and
intermediate moments of inertia in the equator plane. We define
${\bf\hat m}=\be_z$ as a unit vector along the spin axis of the
mantle. The
$xyz$ axes are oriented with respect to the $XYZ$ axes through Euler angles
$\Omega,i,\psi_m$ as shown in Fig. \ref{fig:orbitfixed}, where $\Omega$
is the longitude of the ascending node 
of the equator on the orbit plane, $i$ is the obliquity, and $\psi_m$ is
the angle between the node and the $x$ axis in the equator plane, with
the subscript $m$ again referring to the mantle. 

{\bf [Figure 2]}

The expression for $\vec N$ in spherical coordinates relative to the
body axes (Eq. (\ref{eq:torque})) is written in terms of the $xyz$
components, and these are transformed to the $XYZ$ components through
the usual series of three rotations through the Euler angles. With
the unit vector toward the Sun, ${\bf\hat r}=\cos{f}{\be_{\s
X}}+\sin{f}\be_{\s Y} = \sin{\theta}\cos{\phi}\,\be_x+
\sin{\theta}\sin{\phi}\,\be_y +\cos{\theta}\,\be_{z}$, where
$\be_{X,Y,Z}$ and $\be_{x,y,z}$ are unit vectors along the respective
axes,  scalar products
${\bf\hat r}\cdot \be_x,\be_y,\be_z$ yield the expressions for the
products of circular functions of the spherical coordinates in $\vec N$
in terms of the Euler angles and the true anomaly $f$ when
$\be_{x,y,z}$ are expressed in the $XYZ$ system.  These results and
the transformation yield $\vec N$ in the $XYZ$ system in terms of the
Euler angles  and the true anomaly.
\begin{eqnarray}
\vec N&=&-\frac{GM_{\s\odot}M_{\s M}R^2}{r^3}\bigg\{3J_2\sin{i}\cos{i}
\sin{(f-\Omega)}\big[\sin{f}\,\be_{\s X}-\cos{f}\,\be_{\s
Y}\big]\nonumber\\
&&-1.5J_2\sin{i}\sin{2(f-\Omega)}\,\be_{\s Z}\nonumber\\
&&+6C_{22}\sin{i}[\cos{i}\cos{2\psi}\sin{(f-\Omega)}-
\cos(f-\Omega)\sin{2\psi}](-\sin{f}\be_{\s X}+\cos{f}\be_{\s 
Y}) \nonumber\\
&&+6C_{22}\bigg[-\frac{3+\cos{2i}}{4}\sin{2(f-\Omega)}
\cos{2\psi}+\cos{i}\cos{2(f-\Omega)}\sin{2\psi}\bigg]\,\be_{\s Z}\bigg\}
\label{eq:orbitframetorque} 
\end{eqnarray}
To eliminate high frequency terms due to the orbital motions, we
average Eq. (\ref{eq:orbitframetorque}) over an orbit period keeping
everything constant except $f$ and $\psi_m$.  With $\dot\psi_m=1.5n$
($n=$ mean motion), we can write $\psi_m=\psi_{m0}+1.5M$, where $M$ is the
mean anomaly and $\psi_{m0}$ is constant over the orbit period. With this
substitution for $\psi_m$, we expand the circular functions in
Eq. (\ref{eq:orbitframetorque}) to display products of circular
functions of $f$ and $M$ with $a^3/r^3$. The terms yielding non zero
averages over the orbit period are then
\begin{eqnarray}
\left\langle\frac{a^3}{r^3}\right\rangle&=&\frac{1}{2\pi}\int_0^{2\pi}
\frac{a^3}{r^3}\frac{dM}{df}df=\frac{1}{(1-e^2)^{3/2}}\nonumber\\
\left\langle\frac{a^3}{r^3}\cos{3M}\right\rangle&=&\frac{53e^3}{16}
+\cdots\nonumber\\
\left\langle\frac{a^3}{r^3}\cos{2f}\cos{3M}\right\rangle&=&\frac{1}{2}
\left(\frac{7}{2}e-\frac{123}{16}e^3+\cdots\right)\nonumber\\     
\left\langle\frac{a^3}{r^3}\sin{2f}\sin{3M}\right\rangle&=&\frac{1}{2}
\left(\frac{7}{2}e-\frac{123}{16}e^3+\cdots\right), \label{eq:averages}
\end{eqnarray}
where $r=a(1-e^2)/(1+e\cos{f})$ and $dM/df=r^2/(a^2\sqrt{1-e^2}$) have
been used. Some of these averages are easily done directly, whereas
others are easily evaluated with the help of Cayley's (1859)
tabulated expansions. 
 
When $M=0$, the $x$ axis is nearly aligned with the $X$
axis as a condition of the spin-orbit resonance.  This will be true
for small $i$ if $\psi_{m0}=-\Omega$.  With this substitution, the
averaged torque on the permanent asymmetric figure of Mercury is 
\begin{eqnarray}
\langle{\vec N}\rangle&=&-n^2M_{\s M}R^2\Bigg\{\left[\frac{3J_2
\sin{i}\cos{i}}{2(1-e^2)^{3/2}}+\frac{3C_{22}\sin{i}(1+\cos{i})}{2}
\left(\frac{7}{2}e-\frac{123}{16}e^3\right)\right]\times\nonumber\\ 
&&(\cos{\Omega}\,\be_{\s X}+\sin{\Omega\,\be_{\s Y}})\nonumber\\
&&+\frac{3C_{22}\sin{i}(1+\cos{i})}{2}\frac{53}{16}e^3(-\cos{\Omega}
\,\be_{\s X}+\sin{\Omega}\,\be_{\s Y})\nonumber\\
&&+\frac{3C_{22}\sin{i}(1-\cos{i})}{2}\frac{53}{16}e^3(\cos{3\Omega}
\,\be_{\s X}+\sin{3\Omega}\,\be_{\s Y})\Bigg\}, \label{eq:avetorque1}  
\end{eqnarray}
where the $Z$ components average to zero---the $Z$ component with
coefficient $J_2$ because $\langle(a^3/r^3)\cos{2f}\rangle=\langle
(a^3/r^3)\sin{2f}\rangle=0$, and the $Z$ component with coefficient
$C_{22}$ because $\sin{(2\psi_{m0}+2\Omega)}$ vanishes with
$\psi_{m0}=-\Omega$.  The first bracketed factor shows how the
traditional precession constant, involving only $J_2$, has been
modified by the spin-orbit resonance with the addition of the $C_{22}$
term, which decreases the precession period by nearly 30\% from that
due to the $J_2$ term alone. (For the Earth, only $J_2$ is important
in determining the rate of precession, since terms involving $C_{22}$
for the noncommensurate, rapid spin average to zero, and in addition,
$C_{22}\ll J_2$.)

Before writing $\langle{\vec N}\rangle$ in a form suitable for solving
the equations for the spin vector motion, it is instructive to write
Eq. (\ref{eq:avetorque1}) in a different way. The following identities
are needed.
\begin{eqnarray}
\cos{\Omega}\be_{\s X}-\sin{\Omega}\be_{\s Y}
&=&\cos{2\Omega}(\cos{\Omega}\be_{\s X}+\sin{\Omega}\be_{\s Y})
\nonumber\\
&&+\sin{2\Omega}(\sin{\Omega}\,\be_{\s X}-\cos{\Omega}\,\be_{\s Y})
\nonumber\\
\sin{i}(\cos{\Omega}\be_{\s X}+\sin{\Omega}\be_{\s Y})&=&-{\bf\hat
m}\times \be_{\s Z}\nonumber\\
\sin{i}(\sin{\Omega}\be_{\s X}-\cos{\Omega}\be_{\s Y})
&=&-({\bf\hat m}\times\be_{\s Z})\times\be_{\s Z} \label{eq:identities}
\end{eqnarray}
If we rearrange terms and employ some trigonometric identities along
with Eqs. (\ref{eq:identities}) in Eq. (\ref{eq:avetorque1}), we can
write 
\begin{eqnarray}
\frac{\langle{\vec
N}\rangle}{C_m\dot\psi_m}&=&[K_1\cos{i}+K_3(1- 
\cos{i})]({\bf\hat m}\times\be_{\s Z})\nonumber\\
&&-K_2\cos{i}\cos{2\Omega}({\bf\hat m}\times\be_{\s Z})\nonumber\\
&&-K_2\sin{2\Omega}({\bf\hat m}\times\be_{\s Z})\times\be_{\s Z}.
\label{eq:avetorque3}
\end{eqnarray} 
The constants $K_{1,2,3}$ are defined by
\begin{eqnarray}
K_1&=&\frac{n^2M_{\s M}R^2}{C_m\dot\psi_m}\left[\frac{3J_2}{2(1-e^2)
^{3/2}}+3C_{22}\left(\frac{7}{2}e-\frac{123}{16}e^3\right)\right],
\nonumber\\ 
K_2&=&\frac{3n^2M_{\s M}R^2}{C_m\dot\psi_m}C_{22}\frac{53}{16}e^3,
\nonumber\\
K_3&=&\frac{3n^2M_{\s M}R^2}{2C_m\dot\psi_m}C_{22}\left(\frac{7}{2}e-
\frac{123}{16}e^3\right) 
\label{eq:Kconst}
\end{eqnarray}

The ${\bf\hat m}\times\be_{\s Z}$ terms in Eq. (\ref{eq:avetorque3})
describe the retrograde precession of ${\bf\hat m}$ about the orbit
normal, but now with a rate that varies with twice the precession
frequency. The triple cross product term varies the obliquity $i$ with
the same frequency.  Recall that the longitude of the ascending node
of the equator on the orbit plane $\Omega$ is measured from the perihelion
direction. The precession rate is minimal when $\Omega=0\,{\rm
or}\,\pi$ and maximal when $\Omega=\pm\pi/2$. The former corresponds
to the node of the equator plane being parallel to the line of apsides
such that at the perihelion, the torque due the axial asymmetry of
Mercury ($K_2$ has $C_{22}$ as a factor.) vanishes.  When $\Omega=\pm\pi/2$,
the node is perpendicular to the line of apsides, and at perihelion, the
long axis is tilted the maximum amount relative to the orbit plane.
The torque and hence the averaged precession rate is thus maximal for
this phase, since the biggest torque due to the egg shape of Mercury
comes at the perihelion when the long axis is tipped above or below
the plane of the orbit while still being oriented toward the Sun as
closely as possible because of the spin-orbit resonance.  For
$\Omega=0$ or $\pi$, the opportunity for the Sun to exert a large
torque on the axial asymmetry at perihelion is lost, and so the
average of the torque around the orbit is minimal.

The triple product term in Eq. (\ref{eq:avetorque3})
lies along the projection of the spin vector ${\bf\hat m}$ on the
orbit plane. So for $\pi/2>\Omega>0$ the obliquity is increasing,
whereas it is decreasing for $0>\Omega>-\pi/2$. Hence, the obliquity
has a minimum when the spin is crossing the $X$ axis ($\Omega=\pm
\pi/2$) and a maximum when it crosses the $Y$ axis ($\Omega=0\,{\rm
or}\,\pi$). The precession rate is maximal when the obliquity is
minimal and {\it vice versa}, and the precession path of the spin is
elliptical with the long axis along the $Y$ axis.  The reader will
notice that there is also a component of the triple vector product
torque that is parallel to ${\bf\hat m}$. This component may not be
real because of the approximations involved in arriving at
Eq. (\ref{eq:avetorque1}), but if it is real, there may be a
displacement of the center of physical libration as a function of
precessional phase.

A good approximation to the precessional motion can be obtained by
neglecting the last term in Eq. (\ref{eq:avetorque1}), since it is O$(i^2)$
smaller than the next smallest term. With ${\bf\hat
m}=\sin{i}\sin{\Omega}\,\be_{\s X}-\sin{i} 
\cos{\Omega}\,\be_{\s Y}+\cos{i}\,
\be_{\s Z}$, we can write
Eq. (\ref{eq:avetorque1}) in still another way with a rearrangement of
terms, 
\begin{eqnarray}
\frac{\langle{\vec N}\rangle}{C_m\dot\psi_m}&=&\left[K_1m_{\s Z}+K_3
(1-m_{\s Z})-K_2\frac{(1+m_{\s Z})}{2}\right]m_{\s Y}\,\be_{\s
X}\nonumber\\
&&-\left[K_1m_{\s Z}+K_3(1-m_{\s Z})+\frac{K_2(1+m_{\s Z})}{2}\right]
m_{\s X}\,\be_{\s Y}, \label{eq:avetorque2} 
\end{eqnarray}
where $m_{\s X,Y,Z}$ are the respective components of ${\bf\hat m}$.
Then
\begin{displaymath}
\frac{d{\bf\hat m}}{dt}=\frac{\langle\vec N\rangle}{C_m\dot\psi_m},
\end{displaymath}
where $\dot\psi_m\equiv 1.5n$ is fixed by the spin orbit resonance and
can be removed from the time derivative. From Eq. (\ref{eq:avetorque2})
\begin{eqnarray}
\frac{dm_{\s X}}{dt}&=&\left[K_1m_z+K_3(1-m_{\s Z})-K_2\frac{(1+m_{\s
Z})}{2}\right]m_{\s Y}\nonumber\\
\frac{dm_{\s Y}}{dt}&=&-\left[K_1m_z+K_3(1-m_{\s Z})+K_2
\frac{(1+m_{\s Z})}{2}\right]m_{\s X},\label{eq:dmdtarray}
\end{eqnarray}
where for small obliquity, with $m_{\s Z}\approx 1$, the solution is
\begin{eqnarray}
m_{\s X}&=&m^0_{\s X}\cos{\left[\left(\sqrt{K_1^2-K_2^2}\right)t +
\zeta\right]}\nonumber\\ 
m_{\s Y}&=&\sqrt{\frac{K_1+K_2}{K_1-K_2}}m^0_{\s X}
\sin{\left[\left(\sqrt{K_1^2-K_2^2}\right)t+\zeta\right]},
\label{eq:precsoln} 
\end{eqnarray}
where $m^0_{\s X}$ and $\zeta$ are constants depending on initial
conditions. With $C_m=0.18M_{\s M}R^2$, $\dot\psi=1.5n$, $J_2=6\times
10^{-5}$, and $C_{22}=1\times 10^{-5}$ (Anderson {\it et al.} 1987),
$K_1=1.117\times 10^{-2}{\rm rad/yr}$ and $K_2=8.391\times 10^{-5}{\rm
rad/yr}$, the precession period of Mercury's spin about the orbit
normal (or the Cassini state) is 562 years.  If $C_m=0.16M_{\s M}R^2$, the
precession period would be reduced to 500 years. If the core were
solid and $C=0.33M_{\s M}R^2$, the precession period would be 1031 years.
This compares to a 1066 year period obtained numerically by Rambaux
and Bois (2004).  The ratio of the axes of the ellipse traced out by the
projection of the spin vector on the orbit plane is $m_{\s Y}^
{max}/m_{\s X}^{max}=\sqrt{(K_1+K_2)/(K_1-K_2)}=1.0075$. This solution
is consistent with the deductions about the precession from
Eq. (\ref{eq:avetorque3}) discussed above. 

\subsection{Tidal Torque \label{sec:tidaltorque}}

The tidal torque is given by
\begin{equation}
{\vec T}=\frac{3k_2GM_{\s\odot}^2R^5}{r^6}({\bf\hat r}\cdot{\bf\hat r}_{\s T})
({\bf\hat r}_{\s T}\times {\bf\hat r}), \label{eq:tidaltorque1}
\end{equation}
where $k_2$ is the second degree potential Love number, ${\bf\hat r}$ is a
unit vector toward the Sun, and ${\bf\hat r}_{\s T}$ is a unit vector
toward the tidal maximum, which is the sub-Sun position on Mercury a
short time $\Delta t$ in the past.
\begin{equation}
{\bf\hat r}_{\s T}={\bf\hat r}-\frac{d\bf\hat r}{dt}\Delta t,
\label{eq:rtide}
\end{equation}  
where the time derivative is relative to the body system of coordinates.
Replacement of ${\bf\hat r_{\s T}}$ with Eq. (\ref{eq:rtide}) yields
\begin{equation}
{\vec T}=\frac{3k_2GM_{\s\odot}^2R^5\Delta t}{r^6}{\bf\hat r}\times
{\bf\dot{\hat r}},\label{eq:tidaltorque2}
\end{equation}
where we have set ${\bf\hat r}\cdot{\bf\hat r}_{\s T}=1$. So with
the generic relation for a vector ${\vec D},\,(d{\vec D}/dt)_{body}= 
(d{\vec D}/dt)_{space}-\vec\omega\times{\vec D}$, and with ${\vec
D}\rightarrow {\bf\hat r}=\cos{f}\,\be_{\s X}+\sin{f}\,\be_{\s Y}$ and
$\vec\omega\rightarrow \dot\psi_m(\sin{i}\sin{\Omega}\,\be_{\s X} -
\sin{i}\cos{\Omega}\,\be_{\s Y}+\cos{i}\,\be_{\s Z})+\dot\Omega\be_{\s
Z}$, it is easy to obtain
\begin{eqnarray}
{\bf\hat r}\times{\bf\dot{\hat r}}&=&\dot\psi_m\sin{i}\cos(f-\Omega)
(-\sin{f}\,\be_{\s X}+\cos{f}\,\be_{\s Y})\nonumber\\
&&+(\dot f - \dot\Omega-\dot\psi_m\cos{i})\,\be_{\s Z} 
\label{eq:rxrd2}
\end{eqnarray}
Eq. (\ref{eq:rxrd2}) is substituted into Eq. (\ref{eq:tidaltorque2})
and averaged over the orbit period.  Useful averages are
\begin{eqnarray}
\left\langle\frac{a^6}{r^6}\dot f\right\rangle&=&n\left(1+\frac{15}
{2}e^2+\frac{45}{8}e^4+\frac{5}{16}e^6\right)/(1-e^2)^6=nf_1(e)\nonumber\\
\left\langle\frac{a^6}{r^6}\right\rangle&=&\left(1+3e^2+
\frac{3}{8}e^4)\right)/(1-e^2)^{9/2}=f_2(e)\nonumber\\ 
\left\langle\frac{a^6}{r^6}\cos{2f}\right\rangle&=&\left(\frac{3}{2}e^2
+\frac{1}{4}e^4\right)/(1-e^2)^{9/2}=f_3(e)\nonumber\\
\left\langle\frac{a^6}{r^6}\cos^2{f}\right\rangle&=&\frac{f_2(e)+
f_3(e)}{2}\nonumber\\ 
\left\langle\frac{a^6}{r^6}\sin^2{f}\right\rangle&=&\frac{f_2(e)-
f_3(e)}{2} \label{eq:efunctions}
\end{eqnarray} 
where $\dot f=n(a^2\sqrt{1-e^2})/r^2$ has been used in the first of
Eqs. (\ref{eq:efunctions}).  The averaged tidal torque is thus
\begin{eqnarray}
\langle{\vec T}\rangle&=&\frac{3k_2GM_{\s\odot}^2R^5\Delta t}{a^6}\bigg\{
-\dot\psi_m\sin{i}\sin{\Omega}\left(\frac{f_2(e)-f_3(e)}{2}\right)
\be_{\s X}\nonumber\\ 
&&+\dot\psi_m\sin{i}\cos{\Omega}\left(\frac{f_2(e)+f_3(e)}{2}\right)
\be_{\s Y}\nonumber\\
&&+[nf_1(e)-f_2(e)\dot\psi_m\cos{i}]\,\be_{\s Z}\bigg\},
 \label{eq:avetide}
\end{eqnarray}
where we have neglected $\dot\Omega$ compared to $n$ and $\dot\psi_m$.

The relationship between the dissipation function $Q$ and $\Delta t$
follows from a simple example.
The dissipation parameter $Q$ for an system oscillating at frequency
$\omega$ is defined by ({\it e.g.} Lambeck, 1980 p 14) 
\begin{equation}
\frac{1}{Q}=\frac{\displaystyle\oint\frac{dE}{dt}dt}{2\pi
E^*}=\omega\Delta t,\label{eq:Q2} 
\end{equation}
where the numerator is the energy dissipated during a complete cycle
of oscillation and $E^*$ is the maximum energy stored during the
cycle. For a tidally distorted, nearly spherical body with the
disturbing body in a circular equatorial orbit, a cycle would
consist of half a rotation of the distorted body relative to the body
causing the tide.  For a complex tide generated by a non circular, non
equatorial orbit of the disturbing body, each periodic term in a Fourier
decomposition of the tide would have its own maximum stored energy and
dissipation over a complete cycle of oscillation.

The response of an oscillator with forcing function $F=A^\pr\sin{\omega
t}$ when $\omega\ll \omega_0$, with $\omega_0$ being the lowest
frequency of free oscillation, can be represented by
$x=B^\pr\sin{\omega(t-\Delta t)}$, where $\Delta t$ is the phase lag in the
response due to the dissipation as was assumed above for the tidal
response. The rate at which the forcing function does work is
$dE/dt=F\dot x=A^\pr B^\pr\omega\sin{\omega t}\cos{\omega(t-\Delta t)}$. Then
\begin{equation}
E(t)=\int_{t_1}^tF\dot xdt=A^\pr B^\pr\omega\bigg[\frac{-\cos{(2\omega
t-\omega\Delta t)}}{4\omega}+\frac{\sin\omega\Delta
t}{2}t\bigg]_{t_1}^t.\label{eq:E(t)}
\end{equation}
The first term in Eq. (\ref{eq:E(t)}) is the periodic storage of
energy in the oscillator, and the second is the secular loss of
energy. The maximum energy stored is just twice the coefficient of the
cosine term, $E^*=A^\pr B^\pr/2$, and the energy dissipated during a complete
period of $2\pi/\omega$ is $\Delta E =A^\pr B^\pr\pi\omega\Delta t$ with
$\sin{\omega\Delta t}\approx\omega\Delta t$.  We use the energy stored
as the energy increment above the minimum energy in the first term
since the stored tidal energy will always be an increase over the
relaxed spherical shape of the body. Hence $1/Q=\omega\Delta t$ as
indicated in Eq. (\ref{eq:Q2}).  Since $\Delta t$ is independent of
frequency, $Q$ is inversely proportional to frequency in this model.
For Earth like materials, $Q$ is almost independent of frequency
for a wide range of frequencies above typical tidal frequencies
(Knopoff, 1964), but to avoid a step function reversal in torque when
a tidal frequency changes sign, it is not unlikely that the above
frequency dependence will prevail for the small tidal frequencies. 

$Q=100$ is typical of Earth-like materials (Knopoff, 1964) and
apparently for Phobos generated tides on Mars (Smith and Born, 1976).
For Mars the 
dominant frequency is $\omega=2(\dot\psi-n_{\s P})$, where $\dot\psi$
is Mars rotational angular velocity and $n_{\s P}$ is Phobos' mean
orbital motion. The factor 2 follows from there being two tidal cycles
for each synodic period. With a rotation period of $24^h37^m23^s$ and
an orbital period of 0.319 days, $\omega=3.14\times 10^{-4}$ rad/s
leading to $\Delta t\approx 32$ seconds for $Q=100$. For Mercury, the
fundamental tidal period is the orbit period, since Mercury rotates
$180^\circ$ relative to the Sun for each orbit. Although one might
infer that $Q$ is inversely proportional to frequency near a
particular frequency to smooth the transition from positive to
negative frequency, it cannot be the case that the same proportionality
constant will apply over a very wide range of frequency as indicated
by $Q\approx 100$ nearly independent of frequency between
approximately 10 and $10^6$ cps for Earth materials (Knopoff,
1964). We can write $\Delta t = P/(2\pi Q_0)$, with $Q_0$
being the value appropriate to the 88 day orbital period $P$. 

\subsection{Damping the Obliquity \label{sec:dampobliquity}}
To evaluate the damping of the free precession, or equivalently, the
rate of change of the obliquity $i$ of the mantle due    
to the torques on the right hand side of the first of
Eqs. (\ref{eq:vectoreqns}), we write
$\cos{i}={\bf e}_o\cdot{\bf\hat m}$. (Recall $\be_o=\be_{\s Z}$.) With
$\vec\beta=C_m\dot{\vec\psi}_m=C_m\dot\psi_m{\bf\hat m}$,   
\begin{equation}
\frac{d\,\cos{i}}{dt}=\frac{d}{dt}\left(\frac{{\be}_o\cdot
\vec\beta}{\beta}\right)=\frac{1}{\beta}\left({\bf
 e}_o\cdot\frac{d\vec\beta}{dt}\right)-\frac{1}{\beta^2}({\bf
 e}_o\cdot\vec\beta)\frac{d\beta}{dt}.\label{eq:dcosidt1}
\end{equation} 
If we note that $d\vec\beta/dt={\vec{\cal T}}$, where
$\vec{\cal T}$ is the total torque acting on the mantle, and that
$d\beta/dt={\bf\hat m}\cdot\vec{\cal T}$, Eq. (\ref{eq:dcosidt1})
becomes
\begin{equation}
\frac{d\,\cos{i}}{dt}= [{\be}_o-({\be}_o\cdot{\bf\hat m})
{\bf\hat m}]\cdot \frac{\vec{\cal T}}{\beta}=\sin{i}
\frac{{\be}_\perp\cdot{\vec{\cal T}}}{\beta}, \label{eq:dcosidt2}
\end{equation}
where $\be_\perp$ is a unit vector lying in the equator plane that is
perpendicular to the node of the equator on the orbit plane. 

From Eqs. (\ref{eq:avetorque3}) and (\ref{eq:dcosidt2}),
\begin{equation}
\sin{i}\frac{\be_{\s\perp}\cdot\langle{\vec N}\rangle}{\beta}=
-K_2\sin^2{i}\cos{i}\sin{2\Omega}, \label{eq:eperpdotN}
\end{equation} 
represents the periodic variation in the obliquity during the
precession.  From Eqs. (\ref{eq:avetide}) and with $\be_{\s\perp}=
-\cos{i}\sin{\Omega}\,\be_{\s X}+\cos{i}\cos{\Omega}\,\be_{\s Y}
+\sin{i}\,\be_{\s Z}$,
\begin{equation}
\sin{i}\frac{\be_{\s\perp}\cdot\langle{\vec T}\rangle}{\beta}=
K_4\sin^2{i}\left[nf_1(e)-\frac{\dot\psi_m
\cos{i}}{2}(f_2(e)-f_3(e)\cos{2\Omega})\right],
\end{equation}
where 
\begin{equation}
K_4=\frac{3k_2GM_{\s\odot}^2R^5}{nQ_0C_m\dot\psi_m a^6},\label{eq:K4}
\end{equation}
with $1/nQ_0$ replacing $\Delta t$ as described above. The functions
$f_i(e)$ are defined in Eqs. (\ref{eq:efunctions}) with
$f_1(e)=1.72,\, f_2(e)=1.37$, and $f_3(e)=0.0779$ for $e=0.206$.

We have only the core-mantle interaction term left to evaluate, which
we determine with the analysis of Goldreich and Peale (1970). We
note first that $K_2\ll K_1$ from Eqs. (\ref{eq:Kconst}),
and since the tidal torque and core-mantle interaction are 
small compared to ${\vec N}$, the zero-order solution of
Eqs. (\ref{eq:vectoreqns}) is approximately a uniform retrograde
precession of the spin vector at the rate $-\dot\Omega$, where
$\dot\Omega$ is the magnitude of the motion of the projection
of the spin vector on the orbit plane. This zero order motion of the
mantle is determined from Eq. (\ref{eq:avetorque2})
\begin{equation}
\frac{d{\bf\hat m}}{dt}
\approx K_1({\bf\hat m}\times\be_o)=K_1\sin{i}
(-\cos{\Omega}\,\be_{\s X}-\sin{\Omega}\,\be_{\s Y}),
\label{eq:zeroordertorq} 
\end{equation}
where we have also omitted the $K_3$ term because of the factor
$1-m_{\s Z}\approx 0$.
From Eq. (\ref{eq:zeroordertorq}) 
\begin{equation}
\dot\Omega=K_1 =\frac{2\pi}{562}\;{\rm yr^{-1}},
\label{eq:Omegadot}  
\end{equation}
where the 562 year precession period is discussed above for
$C_m=0.18M_{\s M}R^2$. If we substitute Eq. (\ref{eq:Omegadot}) into 
Eq. (\ref{eq:zeroordertorq}), restore the core-mantle interaction but
continue to ignore the tidal torque, and divide both equations by
$C_m\dot\Omega^2$ to make them dimensionless, we can write 
\begin{eqnarray}
\frac{d\dot{\vec\psi^\pr}_m}{dt^\pr}&=&\dot\psi_m^\pr({\bf\hat
m}\times\be_o)- 
k^\pr(\dot{\vec\psi}_m^\pr-\dot{\vec\psi}_c^\pr)\nonumber\\
\frac{d\dot{\vec\psi}_c^\pr}{dt^\pr}&=&\frac{k^\pr}{C^\pr}
(\dot{\vec\psi}_m^\pr-\dot{\vec\psi}_c^\pr)\label{eq:normeqns}
\end{eqnarray}
where $t^\pr=\dot\Omega t,\,\dot{\vec\psi}_{m,c}^\pr=
\dot{\vec\psi}_{m,c}/\dot\Omega,\,k^\pr=k/(C_m\dot\Omega)$, 
and $C^\pr=C_c/C_m$. With ${\vec\delta}^\pr=
\dot{\vec\psi}_m^\pr-\dot{\vec\psi}_c^\pr$, we can write 
\begin{equation}
\frac{d{\vec\delta}^\pr}{dt^\pr}=\dot\psi_m^\pr({\bf\hat m}
\times\be_o)-k^\pr
\left(\frac{C^\pr+1}{C^\pr}\right){\vec\delta}^\pr. \label{eq:ddeltadt}
\end{equation} 

If we assume $t^\pr=0$ when $\Omega=0$ ({\it i.e.} with the
ascending node of the equator plane on the orbit plane aligned with
the perihelion direction), we can write $\Omega
=-\dot\Omega t=-t^\pr$ and replace $\Omega$ in
the inertial system representation of ${\bf\hat m}\times\be_o$ in
Eq. (\ref{eq:zeroordertorq}) with $-t^\pr$. Then
\begin{eqnarray}
\frac{d\delta_{\s X}^\pr}{dt^\pr}+\alpha\delta_{\s X}^\pr&=&
-\dot\psi_m^\pr\sin{i}\cos{t^\pr},\nonumber\\
\frac{d\delta_{\s Y}^\pr}{dt^\pr}+\alpha\delta_{\s Y}^\pr&=&
\dot\psi_m^\pr\sin{i}\sin{t^\pr},\nonumber\\
\frac{d\delta_{\s Z}^\pr}{dt^\pr}+\alpha\delta_{\s Z}^\pr&=&0,
\label{eq:deltaeqns}
\end{eqnarray}
where $\alpha=k^\pr(C^\pr+1)/C^\pr$ and $\delta_{\s X,Y,Z}^\pr$ are the
components of ${\vec\delta}^\pr$ along the respective coordinate
axes. The solutions of these equations are
\begin{eqnarray}
\delta_{\s X}^\pr&=&A_1\exp{-\alpha t^\pr}+\dot\psi_m^\pr\sin{i}
\left(\frac{-\alpha\cos{t^\pr}-\sin{t^\pr}}{\alpha^2+1}\right),\nonumber\\
\delta_{\s Y}^\pr&=&A_2\exp{-\alpha t^\pr}+\dot\psi_m^\pr\sin{i}
\left(\frac{-\cos{t^\pr}+\alpha\sin{t^\pr}}{\alpha^2+1}\right),\nonumber\\
\delta_{\s Z}^\pr&=&A_3\exp{-\alpha t^\pr}, \label{eq:deltasolns}
\end{eqnarray}
where $A_i$ are constants determined by initial conditions. 

With $\be_\perp=-\cos{i}\sin{\Omega}\,\be_{\s
X}+\cos{i}\cos{\Omega}\,\be_{\s Y}+\sin{i}\,\be_{\s Z}$, and
$\Omega=-t^\pr$, we substitute the steady state solution for
${\vec\delta}^\pr$ into the normalized form of
Eq. (\ref{eq:dcosidt2}), with  
\begin{equation}
\left\langle-\frac{k^\pr\sin{i}}{\dot\psi_m^\pr}{\vec\delta}^\pr
\cdot\be_\perp\right\rangle=\frac{k^\pr\sin^2{i}\cos{i}}{\alpha^2+1},
\label{eq:dcosicoremantle}
\end{equation}
to yield in dimensionless form
\begin{eqnarray}
\frac{d\cos{i}}{dt^\pr}&=&\sin{i}
\left(\frac{K_2}{K_1}\sin{i}\cos{i}\sin{2t^\pr}+\frac{k^\pr\sin{i}\cos{i}}
{\alpha^2+1}\right)\nonumber\\
&&+K_4\sin^2{i}\bigg\{n^\pr f_1(e)-\frac{\dot\psi_m^\pr\cos{i}}{2}
[f_2(e)+f_3(e)\cos{2t^\pr}]\bigg\},\label{eq:dcosidt3}
\end{eqnarray}    
where $n^\pr=n/\dot\Omega$.
 If we average Eq. (\ref{eq:dcosidt3}) 
over a precession period, the circular functions with $t^\pr$ in their
arguments vanish. Next with $d\cos{i}/dt^\pr=-\sin{i}\,di/dt^\pr$ and setting
$\sin{i}\approx i$ and $\cos{i}\approx 1$, we arrive at 
\begin{equation}
\left\langle\frac{di}{dt^\pr}\right\rangle=-i\left[\frac{k^\pr}
{\alpha^2+1}+K_4\left(n^\pr 
f_1(e)-\frac{\dot\psi_m^\pr}{2}f_2(e)\right)\right].\label{eq:didt}
\end{equation}
The time constant
for the decay of the obliquity due to both a core-mantle interaction
and tidal dissipation is the reciprocal of the coefficient of $i$ in
Eq. (\ref{eq:didt}). 

We relate the coupling constant $k^\pr$ to the core viscosity
by equating the time scale for the spin-up of a viscous liquid in
a closed container with the time scale for relaxation of the
differential rotation. From Eqs. (\ref{eq:deltasolns}), the
dimensioned time scale in the transient part of $\dot\delta_i^\pr$ is 
\begin{equation}
\tau_\delta=\frac{1}{\alpha\dot\Omega}=\frac{C^{\s\prime}}
{(1+C^{\s\prime}) k^\pr\dot\Omega} = \frac{R_c}{(\dot\psi_m\nu)
^{1/2}}, \label{eq:taudelta}  
\end{equation}
where the expression on the right is the time scale for a fluid
with kinematic viscosity $\nu$, rotating differentially in a closed
spherical container with radius $R_c$, to become synchronously
rotating with the container at angular velocity $\dot\psi_m$ (Greenspan
and Howard, 1963). Here $\dot\psi_m=1.5n$, $R_c=0.75R$ is Mercury's core
radius  (Siegfried and Solomon, 1987), $\dot\Omega
=(2\pi/562)\,{\rm yr^{-1}}$ and $C^{\s\prime}=C_c/C_m\approx
1$. So
\begin{equation}
k^\pr=\frac{C^{\s\prime}(\dot\psi_m\nu)^{\frac{1}{2}}}
{(1+C^{\s\prime})\dot\Omega R_c} = 8.59\times
10^{-3}\nu^{\frac{1}{2}}, \label{eq:kprime} 
\end{equation}
where $\nu$ is expressed in ${\rm cm^2/sec}$.
 
The kinematic viscosity of the
Earth's core has been estimated from first principles to lie within
the bounds $4.5\times 10^{-3}\ltwid\nu\ltwid 4.1\times 10^{-2}\,{\rm
cm^2/sec}$ by Wijs {\it et al.} (1998). If we choose $\nu=0.01\,{\rm
cm^2/sec}$, then $\alpha=2k^\pr=1.72\times 10^{-3}$, and its square can be
neglected in the denominator of the first term on the right of
Eq. (\ref{eq:didt}).  With $P=88$ days, $R=2439$ km, $C_m=0.18M_{\s
M}R^2$, $M_{\s M}=3.302\times 10^{26}$ g, the time scale for damping
Mercury's free precession becomes
\begin{equation}
\tau_{prec}=\frac{89\,{\rm years}}{8.59\times 10^{-3}\nu^{1/2}+8.08\times
10^{-3}k_2/Q_0}. \label{eq:tauprecess}
\end{equation}
Theoretical values of Mercury's $k_2\approx 0.3 \,{\rm to}\,0.4$
(Spohn {\it et al.} 2001; Van Hoolst and Jacobs, 2003), and
$Q_0\approx 100$ is typical for Mars and Earth (Smith and Born, 1976;
Knopoff, 1964) such that $k_2/Q_0=0.004$ is a plausible value. With
$\nu=0.01\,{\rm cm^2/sec}$, we find 
$\tau_{prec}\approx 1.00\times 10^5$ years under the action of both
tidal dissipation and core mantle interaction, and
$\tau_{prec}=1.04\times 10^5$ years and $\tau_{prec}=2.76\times 10^6$
years for core-mantle interaction and tidal dissipation respectively
if they acted alone. 

The development of Eqs. (\ref{eq:vectoreqns}) for numerical
integration is given in Appendix A. An advantage of a numerical
approach is that the consequences for the damping of large values of
$\nu$, where the approximations used above are not valid, can be
determined. Since, $\nu\gg 0.01$ is unlikely, we shall not show the
damping histories for large $\nu$, but remark that the damping time
scale is a minimum for $\nu\approx 1000\,{\rm cm^2/sec}$ where the
precession amplitude is nearly critically damped.  For larger $\nu$
the coupling is sufficiently strong that the core nearly follows the
mantle as both precess together and the precession period increases
from $562$ years to somewhat over 1000 years appropriate for a solid
planet. The damping time scale for the precession amplitude is
similarly increased from the minimum as $\nu$ is increased. The
Greenspan and Howard spin-up time scale in Eq. (\ref{eq:taudelta})
also must be replaced by a viscous time scale for large $\nu$ in
relating $\nu$ to $k^\pr$. 

\begin{figure}[h]
\plotone{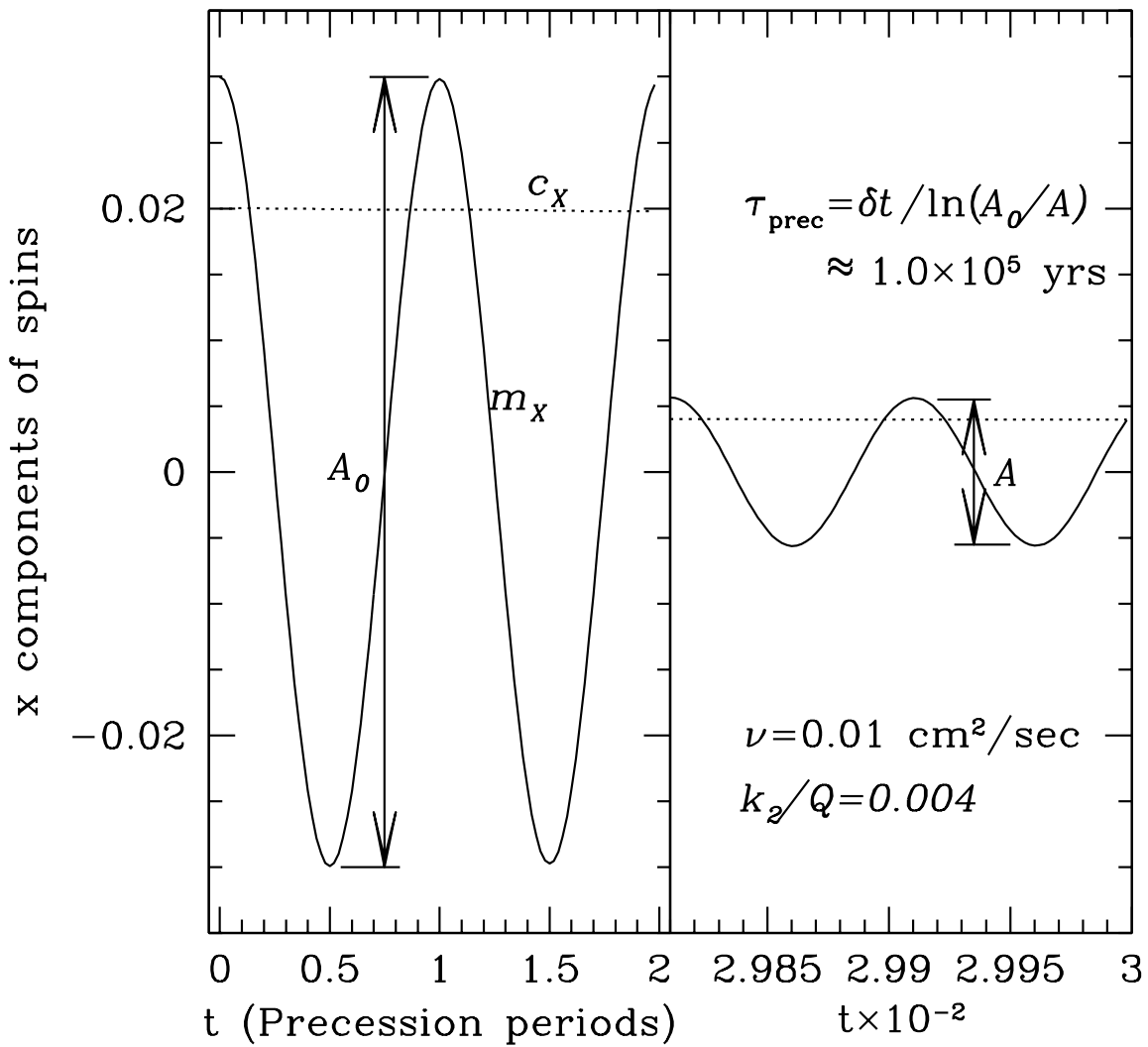}
\caption{Numerically determined damping of Mercury's free precession
with core kinematic viscosity $\nu=0.01\,{\rm cm^2/sec}$ and tidal
factor $k_2/Q_0=0.004$. $m_{\s X}$ and $c_{\s X}$ label the curves for the
$X$ components of the unit vectors along the mantle and core spin axes
respectively. $A_0$ is a small but otherwise arbitrary initial
amplitude.  \label{fig:precdamp}}
\end{figure}

In Fig. \ref{fig:precdamp} we 
demonstrate that the same time scale for damping the precession
amplitude from the combination of core-mantle and tidal dissipation is
obtained numerically as analytically for the same $\nu=0.01\,{\rm
cm^2/sec}$ and $k_2/Q_0=0.004$.  If we set $k=0$, the damping time estimated
numerically similarly to Fig. \ref{fig:precdamp} is  $\approx
2.8\times 10^6$ years compared to $2.76\times 10^6$ years determined
analytically, which confirms the analytic development.  Damping time
scales for reasonable choices of $\nu$ and $k_2/Q_0$ are all short
compared to the solar system age, so we would expect the free
precession to be completely damped, barring excitation by an
unspecified mechanism.  

{\bf [Figure 3]}

\section{Damping of free libration \label{sec:libration}}

\begin{figure}[h]
\plotone{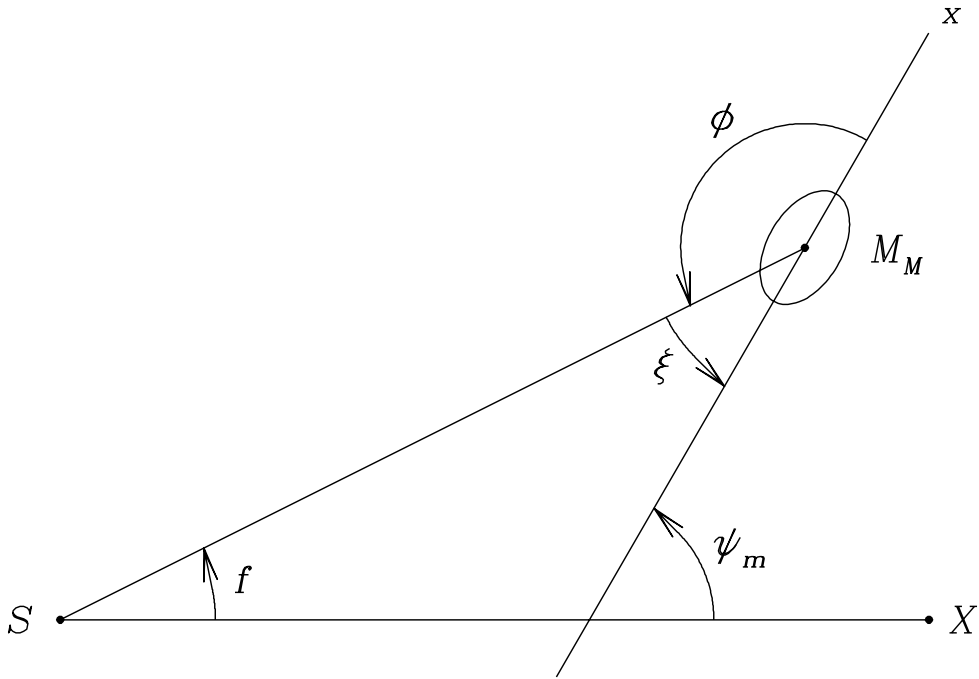}
\caption{Angles used in the discussion of libration in longitude. $SX$
is directed from the Sun to the perihelion of Mercury's orbit and
$M_{\s M}$ denotes Mercury, where $x$ is the axis of minimum moment of
inertia.  \label{fig:libangles}}
\end{figure}

The treatment of the libration damping is considerably simpler than
that of the precession amplitude as we can assume zero obliquity as
well as principal axis rotation. Eqs. (\ref{eq:vectoreqns}) still apply,
but are now scalar equations since there are only torques about the $Z$
axis, which is now coincident with the $z$ axis.  With ${\bf\hat
r}\times \dot{\bf\hat r}=-(\dot\psi_m-\dot f)\be_z$, we can write
\begin{eqnarray}
C_m\ddot\psi_m&=&N+T-k(\dot\psi_m-\dot\psi_c), \nonumber\\
&=&-\frac{3}{2}(B-A)\frac{GM_{\s\odot}}{r^3}\sin{2\xi}-\frac{3k_2GM_{\sun}^2
R^5}{r^6Q_0}\left(\frac{\dot\psi_m-\dot f}{n}\right)-
k(\dot\psi_m-\dot\psi_c),\nonumber\\ 
C_c\ddot\psi_c&=&k(\dot\psi_m-\dot\psi_c), \label{eq:unaveq}
\end{eqnarray}
where $C_{22}=(B-A)/(4M_{\s M}R^2)$ has been used. The relations between
$f,\,\phi,\,\xi,\,{\rm and}\,\psi_m$ are shown in Fig.
\ref{fig:libangles}, where the $X$ axis is now drawn from the Sun to the
perihelion with line $SM$ drawn from the Sun to Mercury.  The angles are
linked to the orbital motion through $\psi_m=\xi+f$, where $f$ is the
true anomaly.

{\bf [Figure 4]}

Mercury's rotation deviates from $\dot\psi_m=1.5n$ by only a
small amount, where $\dot\psi_m$ is the magnitude of
$\dot{\vec\psi_m}$.   We therefore
write $\dot\gamma_m=\dot\psi_m-1.5n$ as the deviation of the spin from
the mean value.  The angular deviation of, say, Mercury's axis of
minimum moment of inertia (``long axis'') from the position it would
have had if the rotation were uniform is then $\gamma_m=\psi_m-1.5M$,
where $M=nt$ is the mean anomaly.  $\gamma_m$ is thus the angle
between Mercury's long axis and the direction to the Sun when Mercury
is at perihelion. The equation of rotational motion for the mantle
becomes 
\begin{equation}
\ddot\gamma_m + \frac{3}{2}n^2\frac{B-A}{C_m}\frac{a^3}{r^3}\sin
{(2\gamma_m+3M-2f)} = \frac{T}{C_m} -
\frac{k}{C_m}(\dot\gamma_m-\dot\gamma_c), \label{eq:gamddot} 
\end{equation}
where $n^2=GM_{\s\odot}/a^3$ is used, with $a$ being the orbit
semimajor axis. The expansion of $\sin{(2\gamma_m+3M-2f)}$ leads to
factors $(a^3/r^3)\sin{2f}$ and $(a^3/r^3)\cos{2f}$, which when
expanded in terms of the mean anomaly $M$ with the help of Cayley's
tables (1859) gives 
\begin{eqnarray}
\ddot\gamma_m+3\frac{B-A}{C_m}G_{201}(e)\gamma_m&=&-\frac{3}{2}\frac{B-A}
{C_m}\bigg[
\left(1-11e^2+\frac{959}{48}e^4+\cdots\right)\sin{t}\nonumber\\
&&-\left(\frac{1}{2}e+\frac{421}{48}e^3+\cdots\right)\sin{2t}+\cdots
\bigg]\nonumber\\
&&+ \frac{T}{C_m n^2}-\frac{k}{C_m n}\left(\frac{\dot\gamma_m}{n} -
\frac{\dot\gamma_c}{n}\right), 
\label{eq:gamddotnorm}
\end{eqnarray}  
where $e$ is the orbital eccentricity,
$G_{201}(e)=7e/2-123e^3/16+\cdots$ is a Kaula (1966) eccentricity
function. We have made the equation dimensionless by dividing by $n^2$
and  letting $nt=M\rightarrow t$. Also,
$\sin{2\gamma_m}\rightarrow 
2\gamma_m$, $\cos{2\gamma_m} \rightarrow 1$ have been used, and a
Fourier series in $t$ multiplied by $\gamma_m$ has been omitted from
the right hand side of the equation, since it is so much smaller than
the unit coefficient of the displayed series.   

If we neglect the weak
tidal and core-mantle torques for the moment, Eq.(\ref{eq:gamddotnorm})
is a simple forced harmonic oscillator equation describing the free
and forced librations of longitude for Mercury.  The amplitude of the
$\sin{2t}$ term is about 11\% of that of the $\sin{t}$ term and higher
order terms are even less.  Since the overall amplitude of the
physical libration is not changed by this second term in the series,
and higher order terms contribute negligibly, we retain only the $\sin
t$ in defining the amplitude of the physical libration in
Eq. (\ref{eq:libamp}). Although this truncation of the series gives a 
reasonably good approximation to the physical libration angle, it
gives a poor representation of the deviation of the angular velocity
from the mean value.  This is shown in Fig. \ref{fig:physlib} where
the flat tops on the angular velocity curve at perihelion result from
the fact that the orbital angular velocity is close to the spin 
angular velocity near perihelion, and with the long axis nearly
aligned with the direction to the Sun during this time, the torque on
the axial asymmetry is nearly zero. A detailed explanation of the
behavior of $d\psi_m/dt$ near the perihelion is given in Appendix B.

{\bf[Figure 5]}

The free libration is treated by averaging Eq. (\ref{eq:gamddotnorm})
over an orbit period such that the periodic terms responsible for the
physical libration vanish. $T$ is replaced by its averaged value $\langle
T\rangle$ and $\dot\gamma_m$ and $\dot\gamma_c$ are uniform on the
orbital time scale. We have already specified the form of the
frequency dependent tidal torque in Eqs. (\ref{eq:unaveq}), where
$\Delta t$ (defined in Eq. (\ref{eq:rtide})) is expressed in terms of
$\dot\psi_m-\dot f$. 
\begin{eqnarray}
\langle T\rangle&=&\frac{1}{2\pi}\int_0^{2\pi}TdM,\nonumber\\
&=&-\frac{3k_2n^4R^5}{2\pi GQ_{\s 0}\sqrt{1-e^2}}\int_0^{2\pi}\frac{a^4}{r^4}
\left(\frac{3}{2}+\frac{\dot\gamma_m}{n}-\frac{a^2}{r^2}\sqrt{1-e^2}
\right)df,\nonumber\\
&=&-F\left(V+\frac{\dot\gamma_m}{n}\right), \label{eq:Tave}
\end{eqnarray}
where we have used the second form of the mantle equation in
Eqs. (\ref{eq:unaveq}), $dM/df=r^2/(a^2\sqrt{1-e^2})$, $\dot
f=na^2\sqrt{1-e^2}/r^2$, and where  
\begin{eqnarray*}
F&=&\frac{3k_2n^4R^5}{GQ_0(1-e^2)^{9/2}}\left(1+3e^2+\frac{3}{8}
e^4\right)=\frac{3k_2n^4R^5}{GQ_0}f_2(e)\\
V&=&\frac{3}{2}-\frac{(1+15e^2/2+45e^4/8+5e^6/16)}{(1-e^2)^{3/2}(1+3e^2+
3e^4/8)}=\frac{3}{2}-\frac{f_1(e)}{f_2(e)},
\end{eqnarray*}
which agrees with Eq. (24) of Goldreich and Peale (1966), except for
the correction of the coefficient of $e^6$ from 3/16 to 5/16.  

We divide  Eq. (\ref{eq:gamddotnorm}) by $x^2 = 3(B-A)G_{201}(e)/C_m$
and further normalize time by $xt=t^\pr$. Time $t\rightarrow 
t^\pr$ is now 
normalized by the free libration frequency instead of the precession
frequency $\dot\Omega$ as in Section \ref{sec:precession}.
The average of Eq. (\ref{eq:gamddotnorm}) then becomes 
\begin{eqnarray}
\frac{d^2\gamma_m^\pr}{d{t^\pr}^2}+\gamma_m^\pr&=&-k^\pr(\dot\gamma_m^\pr-
\dot\gamma_c^\pr)-F^{\pr}(V+x\dot\gamma_m^\pr)\nonumber\\
\frac{d^2\gamma_c^\pr}{d{t^\pr}^2}&=&\frac{k^\pr}{C^\pr}
(\dot\gamma_m^\pr-\dot\gamma_c^\pr),
\label{eq:gamddotave}
\end{eqnarray}
where $F^\pr=F/C_mx^2n^2,\,k^\pr=k/C_mxn,\, C^\pr=C_c/C_m$ and
$\dot\gamma_{m,c}^\pr=\dot\gamma_{m,c}/xn$. Similar to the treatment of the
precession, the fact that the tidal and core-mantle torques 
on the mantle are small means that the mantle will librate almost as
it would if $k$ and $F$ were zero. We follow the same procedure as in
Section \ref{sec:precession} but now in scalar form. The zero order
solution of the first of Eqs. (\ref{eq:gamddotave}) is then
$\gamma_m=\dot\gamma_{m0}^\pr\sin{t^\pr}$ and 
$\dot\gamma_{m}^\pr=\dot\gamma_{m0}^\pr\cos{t^\pr}$, where
$(\gamma_m,\dot\gamma_m)=(0,\dot\gamma_{m0})$ when $t^\pr=0$. Subtracting the
second of Eqs. (\ref{eq:gamddotave}) from the first yields an equation
for $\delta=\dot\gamma_m^\pr-\dot\gamma_c^\pr$.
\begin{equation}
\frac{d\delta}{dt^\pr}+k^{\s\prime}\frac{1+C^\pr}
{C^\pr}\dot\delta = -\dot\gamma_{m0}^\pr\sin{t^\pr} 
-F^{\s\prime}(V+x\dot\gamma_{m0}^\pr\cos{t^\pr}),
\label{eq:deltadot} 
\end{equation}
where we have substituted the zero order solutions for $\gamma_m$ 
and $\dot\gamma_m^\pr$ on the right. Now $F^{\s\prime}=1.5\times
10^{-2}k_2/Q_0$, 
where we have used $x=\sqrt{3(B-A)(7e/2-123e^3/16)/C_m}=0.0262$,
$C_m=0.18M_{\s M}R^2$, inferred from Siegfried and Solomon (1974),
$(B-A)/C_m=3.5\times 10^{-4}$, and Mercury's mass 
$M_{\s M}=3.302\times 10^{26}$ g.  The coefficient of $\cos{t^\pr}$ in
Eq. (\ref{eq:deltadot}) is a factor of $3.9\times 10^{-4}$ less than
the coefficient of $\sin{t^\pr}$ so we can neglect the former term.
We keep the constant term leading to the solution
\begin{equation}
\dot\delta=A_4\exp{(-\alpha t)}+\frac{\dot\gamma_{m0}}{1+\alpha^2}(\cos{t}
-\alpha\sin{t})-\frac{F^\pr V}{\alpha}, \label{eq:deltasolution}
\end{equation}
where $\alpha=k^{\s\prime}(1+C^{\s\prime})/C^{\s\prime}$ and $A_4$ is an
arbitrary constant. 

The steady state solution for $\dot\delta$ is substituted back into
the first of Eqs. (\ref{eq:gamddotave}) to yield
\begin{equation} 
\frac{d^2\gamma_m^\pr}{d{t^\pr}^2} +
\left(\frac{k^{\s\prime}}{1+\alpha^2}+F^{\s\prime}x 
\right)\frac{d\gamma_m^\pr}{dt^\pr} +
\left(1-\frac{k^{\s\prime}\alpha}{1+\alpha^2} 
\right)\gamma_m^\pr=-\frac{F^{\s\prime}V}{1+C^\pr}, \label{eq:gamddotfinal} 
\end{equation}
where we have used the zero order solutions, $\gamma_m^\pr=
\dot\gamma_{m0}^\pr\sin{t^\pr}$ and $\dot\gamma_{m}^\pr=
\dot\gamma_{m0}^\pr\cos{t^\pr}$ in
the steady state solution for $\dot\delta$ and rearranged terms.
Tidal friction displaces the zero point of the libration by
$-F^\pr V/(1+C^\pr)=1.83\times 10^{-3}k_2/Q_0=7.34\times 10^{-6}$
radians or $\sim 1.5$ arcsec for $k_2/Q_0=0.004$
($x=0.0262,\,V=0.244$). A measure of this displacement
would determine $k_2/Q_0$ for Mercury. (Tidal friction attempts to 
slow Mercury's spin to a value that is less than the current $1.5n$,
so at perihelion, the axis of minimum moment of inertia is displaced
from exact alignment  with the Sun ($\gamma_m\equiv 0$) until the
averaged torque on the permanent deformation balances the tidal torque
and keeps the spin at the resonant rate.)     

The time scale $\tau_{lib}$ for damping the free libration amplitude is
twice the inverse of the coefficient of $\dot\gamma_m$ in
Eq. (\ref{eq:gamddotfinal}).  The transient part of the expression for
$\delta$ in Eq. (\ref{eq:deltasolution}) is the same as that for the
precession analysis, so the coupling constant $k^\pr$ is related to
the kinematic viscosity of the core by Eq. (\ref{eq:kprime}), except
$\dot\Omega$ must be replaced by $xn$. Then 
\begin{equation}
k^\pr=\frac{C^{\s\prime}(\dot\psi\nu)^{\frac{1}{2}}}
{(1+C^{\s\prime})xnR_c} = 1.41\times
10^{-4}\nu^{\frac{1}{2}}. \label{eq:kprime2} 
\end{equation} 
The dimensioned time scale for damping the free libration is given by 
\begin{eqnarray}
\tau_{lib}&=&\frac{2}{(k^{\s\prime}/(1+\alpha^2)+F^{\s\prime}x)xn}\nonumber\\ 
&\approx& \frac{2}{(k^{\s\prime} + F^{\s\prime}x)xn}\nonumber\\
&\approx&\frac{2.92\,{rm years}}{(1.41\times
10^{-4}\nu^{1/2} + 3.93\times 10^{-4}k_2/Q_0)}
\label{eq:taulibration}
\end{eqnarray}
from Eq. (\ref{eq:gamddotfinal}), where $\alpha=2k^{\s\prime}$ is
small and its square can be neglected.  We again choose
$\nu=0.01\,{\rm cm^2/sec}$ after Wijs {\it et al.} (1998)
along with $k_2/Q_0=0.004$ discussed above, we find
$\tau_{lib}\approx 1.86\times 10^5$ years under the action of both tidal
dissipation and core mantle interaction, and $\tau_{lib}=2.07\times
10^5$ years and $\tau_{lib}=1.86\times 10^6$ years for core-mantle
interaction and tidal dissipation respectively if they acted alone.
Eqs. (\ref{eq:unaveq}) are numerically integrated in Fig.
\ref{fig:libdamp} for $\nu=0.01\,{\rm cm^2/sec}$ and $k_2/Q_0=0.004$,
where the short period fluctuations in Fig. \ref{fig:libdamp} are the
88 day forced librations superposed on the free libration.
The approximately determined time scale is $1.8\times 10^5$ years,
which is close to the analytic value of $1.86\times 10^5$
years. Damping time scales for reasonable choices of $\nu$ and
$k_2/Q_0$ are all short compared to the solar system age, so we would
expect the free librations to be completely damped, barring excitation
by an unspecified mechanism. 

{\bf [Figure 6]}

\section{Consequences of Impact \label{sec:impact}}
This is an exercise to determine what size impactor would be necessary
to generate an observable free libration or free precession.
We shall assume Mercury has no free motions prior to the impact, and
that there is a complete transfer to Mercury of the entire angular
momentum of the impactor relative to the center of mass of Mercury. If
the impact ejecta is distributed symmetrically about the point of
impact, it will carry away little angular momentum.  This approximate
conservation of angular momentum allows us to infer that the angular
increment in the amplitude of the precession is simply the component
of the impactor angular momentum that is perpendicular to the spin
axis divided by the spin angular momentum $C_m\dot\psi_m$. The
increment in the free libration amplitude follows easily from the
parallel component of the impactor angular momentum.  We 
shall assume that a free libration amplitude or free precession
amplitude of 0.1 arcmin is detectable. 

Generally, an impactor, which
can be either a comet or asteroid, will strike at a point with
spherical coordinates $R,\theta,\phi$ relative to the body principal
axis system with velocity relative to the center of mass of Mercury of
${\vec V}=V_{\s R}\be_{\s R}+V_\theta\be_\theta+V_\phi\be_\phi$. 
The angular momentum of the impactor relative to the center of mass of
Mercury is given by
\begin{equation}
{\vec L}_{\s I}={\vec R}\times{\cal M}{\vec V}={\cal M}R(V_\theta\be_\phi
-V_\phi\be_{\theta}), \label{eq:impacttorque}
\end{equation}
where ${\cal M}$ is the mass of the impactor. 
With $\be_{\phi}=-\sin{\phi}\,\be_x+\cos{\phi}\,\be_y$ and
$\be_\theta=\cos{\theta}\cos{\phi}\,\be_x+\cos{\theta}\sin{\phi}\be_y-
\sin{\theta}\,\be_z$, we can write 
\begin{eqnarray}
{\vec L}_{\s I}&=&{\cal M}R[(-V_\theta\sin{\phi}-V_\phi\cos{\theta}
\cos{\phi})\be_x\nonumber\\
&&+(V_\theta\cos{\phi}-V_\phi\cos{\theta}\sin{\phi})\be_y\nonumber\\
&&+V_\phi\sin{\theta}\,\be_z], \label{eq:impacttorque2}
\end{eqnarray}
With the increment in
Mercury's angular momentum, $\Delta{\vec L}_{\s M}={\vec L}_{\s I}$, the
components of ${\vec L}_{\s M}={\vec L}_{\s M}^0+\Delta{\vec L}_{\s M}$
along $\be_x$ and $\be_y$ indicates that a free wobble has been
induced. But that is not all, since the redistribution of mass due to
the crater and its ejecta redefines the inertia tensor, so $x,y,z$ are
no longer the principal axes. The new axis of maximum moment being
offset from the original $z$ axis also contributes to the wobble. The
component $\Delta L_z$ means $\dot\psi_m$ has been incremented thereby
creating a free libration in longitude. Transforming ${\vec L}_{\s M}$
to the inertial system defines a new direction of the total spin
angular momentum which no longer coincides with the Cassini state,
which has itself been changed because of the change in the inertia
tensor. ${\vec L}_{\s M}$ now precesses around the new position of the
Cassini state. A random impact thus excites all three free motions
(see Peale, 1975 for a more detailed treatment), but how large must
the impact be for detectable amplitudes?  

First we note that 
\begin{eqnarray}
L_{\s I\parallel}&=&{\cal M}RV_\phi\sin{\theta}\nonumber\\
L_{\s I\perp}&=&{\cal M}R(V_\theta^2+V_\phi^2\cos^2{\theta})^{1/2}
\label{eq:Lparperp}
\end{eqnarray}
are the components of impactor angular momentum just before impact at
coordinates ($\theta,\phi$) that are parallel and perpendicular to the
spin axis.
We consider first the excitation of a free libration, which involves
only $L_{\s I\parallel}$.  The free libration is characterized by
Eq. (\ref{eq:gamddotnorm}), where we neglect the right hand side to
yield a small amplitude pendulum equation, whose frequency of
oscillation is $xn=n\sqrt{3(B-A)G_{201}(e)/C_m}$. Recall that 
$\gamma_m=\psi_m-1.5M$ is the angular deviation of the $x$ axis from the
position it would have had if Mercury were rotating uniformly at
angular velocity $1.5n$. If $\gamma_m=0$ and
$\dot\gamma_m=\dot\gamma_m^0$ at $t=0$, the amplitude of the libration is 
\begin{equation}
\gamma_m^{max}=\dot\gamma_m^0/\omega_0=\frac{L_{\s I\parallel}}
{C_m xn}, \label{eq:gammamax}
\end{equation}
where the final form assumes that all of the libration amplitude is
induced by the impact. 

If we set $\gamma_m^{max}=0.1^\pr= 2.91\times 10^{-5}\,{\rm rad}$ as the
minimum observable, choose $V_\phi=60\,{\rm km/sec}$ as a common
impact velocity, set $C_m=0.18M_{\s M}R^2$ as before, choose
$(B-A)/C_m=3.5\times 10^{-4}$ so that $xn=0.026n$, and maximize
the effect of the collision for a given ${\cal M}$ by assuming
$\theta=\pi/2$, we find the necessary ${\cal M}=1.51\times
10^{15}\,{\rm g}$. For a density of 1 ${\rm g/cm^3}$, the radius of the
impactor is 0.712 km.

Holsapple (1993) has described a careful development of scaling laws
that predict the properties of impact craters from the size and
velocity  of the impactor and the strength and gravitational
acceleration of the body on which the impact occurs. His Eq. (22b)
gives the rim radius for large craters where gravity dominates
the strength of the materials in controlling the outcome of the crater
formation.  
\begin{equation}
R_r=10.14{G^\pr}^{-0.17}{a^\pr}^{0.83}U^{0.34}, \label{eq:rimradius}
\end{equation}
where $R_r$ is the rim radius of the crater just after it is formed in
meters, $G^\pr$ is the ratio of the gravitational acceleration at the
surface of the body to that of the Earth, $a^\pr$ is the radius of the
impactor in meters, and $U$ is the velocity in km/sec.
Eq. (\ref{eq:rimradius}) ignores the ratio of the impactor density
and surface material density raised to a small fractional power. With
Mercury's gravitational acceleration of 363 cm/sec compared to Earth's 980
cm/sec, $a^\pr=712\,{\rm m}$ and $U=V_\phi=60\,{\rm km/sec}$, 
we find $R_r$=11.3
km. For a crater this large, it gets even larger because of slumping of
the walls, with the expression for the final crater size in terms of
$R_r$ given by Eq. (28a) of Holsapple (1993). This is called a
transition from a simple bowl-shaped crater to a complex crater with a
shallow floor and central peak. 
\begin{equation}
\frac{R_f}{R_r}=1.02\left(\frac{R_f}{R_*}\right)^{0.079},
\label{eq:finalradius} 
\end{equation}
where $R_f$ is the final rim radius, and $R_*$ is the transition
radius from simple to complex craters. $R_*$ is a function of the
surface gravity with $R_*=5$ km for Mercury (Pike, 1988) (9.5 km for
the Moon and 1.5 km 
for Earth and Venus). Substitution of this value of $R_*$ into
Eq. (\ref{eq:finalradius}) yields 12.4 km as the rim radius expected
for the impact event leading to a $0.1^\pr$ free libration amplitude for
Mercury. This is a minimum value because of the choice of the extreme
optimum values of $V_{\s \phi}$ and $\sin\theta$ in
Eq. (\ref{eq:gammamax}). It should be noted that the Holsapple crater
radii are for vertical 
impacts, whereas we have calculated the angular momentum impulse for
an almost grazing impact. However, this caveat does not change the
conclusion that even the least detectable amplitude of a free
libration of $0.1^\pr$ excited by an impact would leave a more than 20 km
diameter pristine scar on the surface of Mercury.

For an impact excitation of the precession, the angular amplitude
would be 
\begin{equation}
\Delta i=\frac{L_{\s I\perp}}{C_m\dot\psi_m}=\frac{{\cal M}V_\theta R}
{C_m\dot\psi_m}, \label{eq:Deltai}
\end{equation}
where we have assumed the impact is along a meridian for simplicity.
If $\Delta i=0.1^\pr$, $V_\theta=60\,{\rm km/sec}$, and $C_m=0.18M_{\s
M}R^2$, then ${\cal M}=8.72\times 10^{16}\,{\rm g}$. The impactor radius is
then $a^\pr=2.75\,{\rm km}$ for a density of $1\,{\rm g/cm^3}$. It
takes a much more energetic impact to 
excite an observable free precession than an observable free
libration. These numbers, when substituted into
Eqs. (\ref{eq:rimradius}) and (\ref{eq:finalradius}), yield
$R_f=41.7\,{\rm km}$, and the pristine crater left behind from an
impact causing a $0.1^\pr$ increment in the amplitude of the free
precession would be more than 80 km in diameter. The fact that the
Cassini state itself has shifted slightly because of the redefinition
of the inertia tensor would not change the free precession amplitude
sufficiently to alter this conclusion.

In Table 1 of Levison and Duncan (1997), the rate of impact of Jupiter
family comets on Mercury is given as $8.6\times 10^{-10}$
comets/year. These are for comets of absolute magnitude $< 9$, which
correspond to radii $a^\pr\gtwid 1\,{\rm km}$ (H. Levison, private
communication 2005).  The collision rate of very long period comets is
probably small compared to this rate for the Jupiter family
comets. (H. Levison, private communication 2005). More than a billion
years would elapse on average between impacts on Mercury of sufficient
size to excite a $0.1^\pr$ amplitude of the free libration in longitude,
and because the larger impactor required, even more time between
excitations of a $0.1^\pr$ amplitude of the free precession.  Comparing
this time between excitations with the O$(10^5)$ year damping times for
the libration and precession amplitudes means collisions almost
certainly could not be the cause of any observable free libration or
free precession.  

\section{Summary and Conclusions \label{sec:summary}}
The spin orbit resonance of Mercury leads to a 30\% reduction of the
period of precession of the spin about the Cassini state from a
contribution of the $C_{22}$ term, compared to the usually derived
precession period depending on $J_2$ alone.  This effect also leads to
an elliptical precessional path of the spin, where the ratio of the
axes of the ellipse is about 1.0075.  There is a slight variation in
the precession rate about the elliptical path with the maximum
occurring when the separation of the spin from the Cassini state is
minimal. 

The simultaneous effects of tidal dissipation and a viscous
liquid core-solid mantle coupling lead to time scales for damping
Mercury's free libration in longitude and free precession of the spin
about Cassini state 1 that are near $10^5$ years. Because Mercury's liquid
core is so large, the time scale for damping the precession amplitude
by the core-mantle coupling acting alone is a factor of about 25
smaller than the time scale due to the tides alone for plausible
choices of the parameters. For the libration damping, the core-mantle
time scale is a factor of about 10 smaller than that for tidal
friction.  For all reasonable choices of the parameters, the time
scales are short compared with the age of the solar system. So if a
free libration or free precession amplitude is found by the radar
experiments or by the observations of the MESSENGER or BepiColombo
spacecraft, still unspecified recent or ongoing excitation mechanisms
must be sought. It is possible that there exist observable forced
deviations of the spin axis from the Cassini state. If the forcing
terms are identified, the amplitudes and phases of such deviations
should be known and should not hinder the identification of the
Cassini state obliquity.  On the other hand, the forcing terms are
likely to be of fairly long period, as is likely if the amplitudes of
the deviations from the Cassini state are observable. As such their
identity may not be obvious and the Cassini state position could be
thereby uncertain until careful numerical integrations mapped such
forced motions.

 An observable 
libration amplitude (assumed to be 
$0.1^\pr$) is much easier to generate by impact than an observable
precession. But such impact excitation of either motion is very unlikely,
since there is an estimated  average time span between impacts of
sufficient size of about $10^9$ years with current cometary fluxes.
Excitation by impact no more than a few damping time scales in the
past would leave a fresh crater larger than 20 km diameter for the
excitation of an observable amplitude of libration and larger than 80
km diameter for the excitation of an observable precession amplitude.  
An observable free precession will make the obliquity of the Cassini
state, and thereby the determination of $C/M_{\s M}R^2$ uncertain. But an
observable free libration does not hamper the determination of the
physical libration amplitude. If Mercury's core were solid, the term
with the kinematic 
viscosity would vanish in the expressions for the damping times of the
precession and libration (Eqs. (\ref{eq:tauprecess}) and
(\ref{eq:taulibration}). But the resulting damping times due to the tides
alone would be longer that these equations would indicate for $\nu=0$,
because the constants would then involve the moment of inertia of the
entire planet instead of just that of the mantle.

This analysis applies to other bodies in spin-orbit resonances
such as the Moon, the regular satellites of the major planets, and
Pluto's satellite Charon. The terms selected from the expansion of the
torque expressions by the averaging process would correspond to
rotation synchronous with the the orbital mean motions instead of 1.5
times the mean motion used here. There are similar contributions to
the rate of precession about the Cassini state from $C_{22}$, since
the synchronous rotation ensures terms involving $C_{22}$ do not
average to zero and $C_{22}$ will be comparable to $J_2$ in magnitude. 
There are indications that some of these bodies even have liquid
layers in their interiors ({\it e.g.}, Spohn and Schubert, 2003),
although Europa would be a special case with a liquid ocean ({\it e.g.}
Greenberg, 2004)

It will be more fun to seek alternative dynamic mechanisms of 
excitation of observed free motions of Mercury than if no such motions are
found.  A possible albeit unlikely excitation of libration may
lie in a coupling between the torques on the permanent deformation as
affected by the spin-orbit resonance and the small obliquity of the
Cassini state.  Although such a libration would not be ``free,'' a
near resonant forcing frequency could mimic a free libration.
The investigation of such speculation is a future objective. 

\section{Appendix A: Numerical Integration of the Precession Damping}
We want to integrate Eqs. (\ref{eq:vectoreqns}) over the precession
time scale without following high frequency motions on the orbit or
libration time scales. The averaged torque on the permanent
deformation $\langle{\vec N}\rangle$ is given in
Eq. (\ref{eq:avetorque1}),  where we set $\psi_m=\psi_0+1.5M$ with
$\psi_0=-\Omega$ to accommodate the conditions of the spin-orbit
resonance in the averaging process. The last term of
Eq. (\ref{eq:avetorque1}) is of order $i^2$ smaller than the next
smallest term, and it is neglected in Eq. (\ref{eq:avetorque2})
written in terms of the components of the unit vector along the spin
axis ${\bf\hat m}$. It is this latter form of the averaged torque on
the permanent deformation that we will use here.

The averaged tidal torque is given by Eq. (\ref{eq:avetide}).
The component of $\langle{\vec T}\rangle$ parallel to the spin axis is
balanced by a slight offset of the axis of minimum moment of inertia
from the direction to the Sun at perihelion. However, we have removed
that degree of freedom in $\langle{\vec N}\rangle$ by assigning
$\psi_0=-\Omega$ after Eq. (\ref{eq:avetorque1}). In order to avoid
a secular change in $\dot\psi_m$, we keep only the component
of $\langle{\vec T}\rangle$ perpendicular both to the spin vector and
to the line of nodes.  The unit vector in this direction is
$\be_{\s\perp}$ defined in Eq. (\ref{eq:dcosidt2}). From
this last equation, we see that 
this is the only component of $\langle{\vec T}\rangle$ that affects the
obliquity.  With $\psi_0=-\Omega$ and the elimination of the tidal
acceleration of the spin, we have eliminated integration over the free
libration period leaving only the precession time scale in the
calculation.  So in place of $\langle{\vec T}\rangle$, we shall use 
\begin{eqnarray}
\frac{\langle{\vec T}\rangle\cdot\be_{\s\perp}}{C_m\dot\psi_m}
\be_{\s\perp}&=&K_4\sin{i}\left[nf_1(e)-\frac{\dot\psi_m
\cos{i}}{2}(f_2(e)-f_3(e)\cos{2\Omega})\right]\times\nonumber\\
&&(-\cos{i}\sin{\Omega}\,\be_{\s X}+\cos{i}\cos{\Omega}\,\be_{\s Y}
+\sin{i}\,\be_{\s Z})\label{eq:tidedoteperp}
\end{eqnarray}  
where $K_4$ is defined by Eq. (\ref{eq:K4}) with
$1/(nQ_0)$ replacing $\Delta t$ as described in Section
\ref{sec:tidaltorque}.  Inclusion of the $f_3(e)$ part of the 
bracketed coefficient in Eq. (\ref{eq:tidedoteperp}) leads to a
$\sin{i}$ in the denominator, which complicates the calculation.
However, $f_3(e)\approx 0.056f_2(e)$ for $e=0.206$ so its omission
will lead to a small error in the motion of the spin axis and an
indiscernible error in the damping time scale, With this additional
approximation, we arrive at 
\begin{equation}
\frac{\langle{\vec T}\rangle\cdot\be_{\s\perp}}{C_m\dot\psi_m}
\be_{\s\perp}=K_4[nf_1(e)-\dot\psi_mm_{\s Z}f_2(e)][-m_{\s Z}m_{\s
X}\,\be_{\s X}-m_{\s Z}m_{\s Y}\,\be_{\s Y}+(m_{\s X}^2+m_{\s
Y}^2)\,\be_{\s Z}],\label{eq:avetide2}
\end{equation}
where $m_{\s X,Y,Z}$ are the components of the unit vector along the
spin axis of the mantle, $\bf\hat m$.  With $d
(C_m\dot{\vec\psi}_m)/dt=C_m\dot\psi_m d{\bf\hat m}/dt$ where
$\dot\psi_m=3n/2={\rm constant}$, we can average Eq. (\ref{eq:vectoreqns})
over an orbit period, insert the averaged expressions $\langle{\vec
N}\rangle$ and the component of $\langle{\vec T}\rangle$ parallel
$\be_{\s\perp}$ from Eqs. (\ref{eq:avetorque2}) and
(\ref{eq:avetide2}), divide by $K_1$ to make the equations
dimensionless, set $K_1t=t^\pr$ and $k/(C_mK_1)=k^\pr$, and write
six scalar equations in six unknowns, $m_{\s X,Y,Z},c_{\s X,Y,Z}$,
where $c_{\s X,Y,Z}$ are the components of the unit vector parallel to
the spin of the core.
\begin{eqnarray}
\frac{dm_x}{dt^\pr}&=&\left[m_{\s Z}+\frac{K_3}{K_1}(1-m_{\s Z})-
\frac{K_2}{K_1}\frac{1+m_{\s Z}}{2}\right]m_{\s Y}\nonumber\\
&&-\frac{K_4n}{K_1}\left[f_1(e)-\frac{3}{4}f_2(e)m_{\s Z}\right]m_{\s
Z}m_{\s X}-k^\pr(m_{\s X}-c_{\s X}),\nonumber\\
\frac{dm_y}{dt^\pr}&=&-\left[m_{\s Z}+\frac{K_3}{K_1}(1-m_{\s Z})+
\frac{K_2}{K_1}\frac{1+m_{\s Z}}{2}\right]m_{\s X}\nonumber\\
&&-\frac{K_4n}{K_1}\left[f_1(e)-\frac{3}{4}f_2(e)m_{\s Z}\right]m_{\s
Z}m_{\s Y}-k^\pr(m_{\s Y}-c_{\s Y}),\nonumber\\
\frac{dm_z}{dt^\pr}&=&\frac{K_4n}{K_1}\left[f_1(e)-\frac{3}{4}f_2(e)m_{\s
Z}\right](m_{\s X}^2+m_{\s Y}^2)-k^\pr(m_{\s Z}-c_{\s Z}),\nonumber\\
\frac{dc_{\s X}}{dt^\pr}&=&\frac{k^\pr}{C^\pr}(m_{\s X}-c_{\s X}),\nonumber\\
\frac{dc_{\s Y}}{dt^\pr}&=&\frac{k^\pr}{C^\pr}(m_{\s Y}-c_{\s Y}),\nonumber\\
\frac{dc_{\s Z}}{dt^\pr}&=&\frac{k^\pr}{C^\pr}(m_{\s Z}-c_{\s Z}) 
\label{eq:numericaleqs}
\end{eqnarray}
are the equations that are integrated numerically and that describe the
damping of the precession amplitude due to a core-mantle interaction
and tidal dissipation. Here $C^\pr=C_c/C_m$, and we have set
$\dot\psi_m=1.5n$ in the term from the tides.  

\section{Appendix B: Description of torque reversal near
perihelion.}
\begin{figure}[h]
\plotone{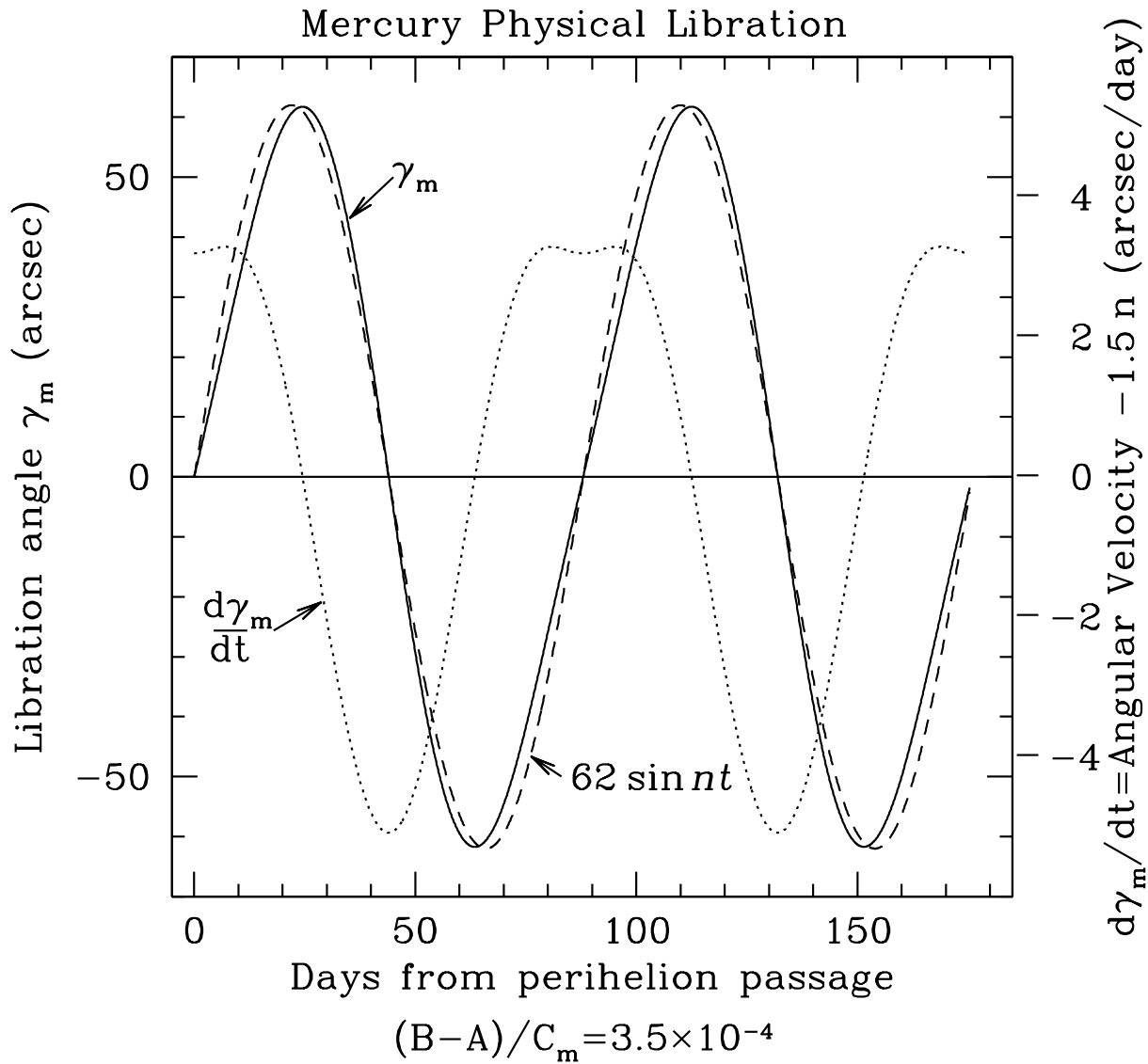}
\caption{Forced libration of Mercury showing the deviations of the
motions from pure harmonic form.  The flat portion of $d\gamma_m/dt$ results
from the fact that the rotation is nearly synchronous with the orbital
motion for a range of true anomaly $f$ on either side of the
perihelion while the axis of minimum moment of inertia is nearly
pointing to the Sun. (See Appendix B.)
\label{fig:physlib}}
\end{figure}

\begin{figure}[h]
\plotone{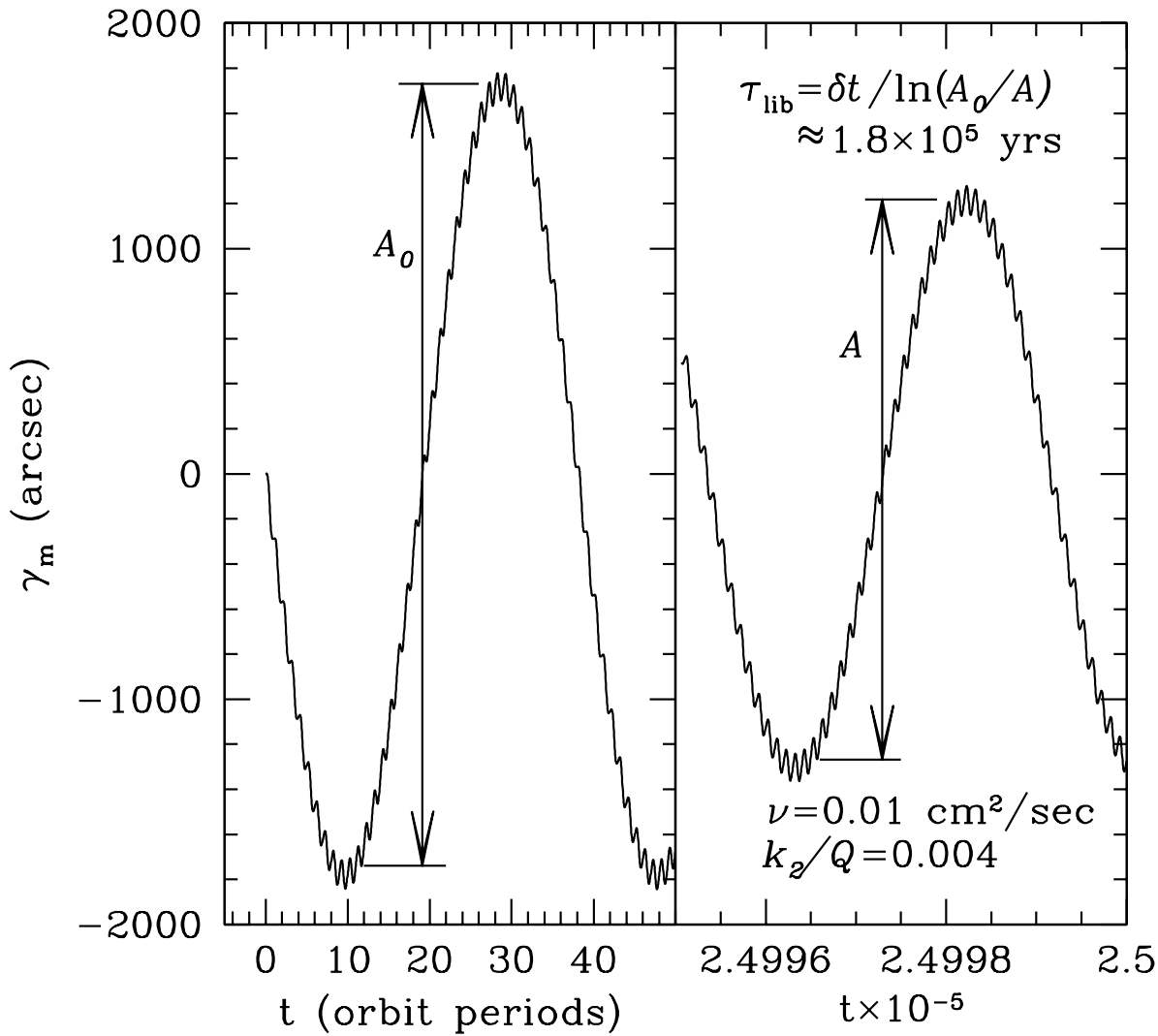}
\caption{Numerically determined damping of Mercury's
free libration in longitude with core kinematic viscosity
$\nu=0.01\,{\rm cm^2/sec}$ and tidal factor $k_2/Q_0=0.004$. The high
frequency oscillations superposed on the free libration are the 88 day
forced librations. $(B-A)/C_m=3.5\times 10^{-4}$ determines the free
libration period of 9.2 years. $A_0$ is a small but otherwise arbitrary
initial amplitude. \label{fig:libdamp}} 
\end{figure}

This is an explanation of the nearly flat top on $d\gamma_m/dt$ in
Fig. \ref{fig:physlib}. In this figure there is a
local minimum in $\dot\gamma_m$ at perihelion and two local maxima at
equal times before and after perihelion.  The angular acceleration is
zero at these extremes and corresponds to the axis of minimum moment
of inertia (long axis) of Mercury being aligned with the direction to
the Sun.  We ignore the slight variation in rotation due to the physical
librations in determining the torque on Mercury by assuming it is
rotating uniformly with $\dot\psi_m=1.5n$.  At
perihelion, the long axis is also pointing toward the Sun, but $\dot
f>\dot\psi_m=1.5n$ at this point in the orbit. The long axis does not
keep up with the motion of the Sun in Mercury's sky as Mercury passes
perihelion, so it starts to point away from the Sun in the direction
of the orbital motion. The angle between the Sun and the long axis
continues to grow and reaches a local maximum when $\dot
f=\dot\psi_m$, which occurs when $f= 25.743^\circ$ and
$M=16.777^\circ$, 4.101 days after perihelion passage. The torque on
Mercury during this time tends to increase the angular velocity
corresponding to the positive slope on $d\dot\gamma_m/dt$ in Fig.
\ref{fig:physlib} just after perihelion passage. The increasing slope
has an inflection at and starts to decrease after the time when $\dot
f=\dot\psi_m$, which is when the angle between the long axis and the
Sun starts to decrease. The slope of the $d\psi_m/dt=d\gamma_m/dt$
curve reaches zero when the axes are again aligned. To bring the long
axis back into 
alignment with the Sun after perihelion passage, Mercury must rotate
through the same angle as it has moved in its orbit, so that $f=1.5M$
at the time of the first post perihelion alignment.  This corresponds
to $f=44.442^\circ,\,M=29.628^\circ$ at 7.242 days after perihelion
passage. After this time, the long axis points on the other side of
the Sun and the angular velocity starts to decrease.  

The motion is symmetric with respect to the perihelion passage so the
numbers corresponding to the approach to perihelion are the negatives
of those above.  It is ascertained that Mercury is almost
synchronously rotating with its instantaneous orbital motion near the
perihelion, during which time the long axis never deviates much from
pointing toward the Sun.  Hence, $\dot\gamma_m$ is almost constant for
a time near the perihelion as shown in Fig. \ref{fig:physlib}, with
the small deviations from a constant value explained above.

\section{Acknowledgements}

Thanks are due H. Levison for directing me to his results for the rate
of cometary impacts on Mercury, to R. Strom for information on crater
morphologies. and to J. Wisdom for reminding me of two relevant papers of
my own. Jacques Henrard and an anonymous referee provided useful
suggestions that improved the manuscript.  This work is supported in
part by the Planetary Geology and Geophysics Division of NASA under
grant NAG5 11666 and by the MESSENGER mission to Mercury.

\section{References \label{sec:references}}
\parindent=0pt
Anderson, J.D., Colombo, G., Esposito, P.B., Lau, E.L., Trager, T.B.
1987. The mass, gravity field and ephemeris of Mercury. Icarus
71, 337-349. 

Anselmi, A., Scoon, G.E.N. 2001. BepiColombo, ESA's Mercury
Cornerstone mission. Planet. Space Sci. 49, 1409-1420. 
 
Cayley, A. 1859. Tables of the developments of functions in the theory of 
elliptic motion, Mem. Roy. Astron. Soc. 29, 191-306.

Colombo, G.  1966. Cassini's second and third laws.  Astron. J.
71, 891-896.

Goldreich, P., Peale, S.J. 1966. Spin orbit coupling in the solar
system.  Astron. J.  71, 425-438.

Goldreich, P., Peale, S.J.  1970. The obliquity of Venus, 
Astron. J.  75, 273-284.

Greenberg, R. 2004. Europa The Ocean Moon: Search For An Alien
Biosphere, Springer, Berlin, (Praxis, Chichester, UK). 

Greenspan, H.P., Howard, G.N.  1963. On the time dependent motion
of a rotating fluid,  J. Fluid Mech.  17, 385-404.

Harder, H., Schubert, G.  2001. Sulfur in Mercury's core? 
Icarus  151, 118-122.

Holin, I.V.  1988.  Izvestiya Vysshikh Uchebnykh Zavedenii,
Radiofizika  31, 515.

Holin, I.V.  1992.  Izvestiya Vysshikh Uchebnykh Zavedenii,
Radiofizika  35, 433.

Holin, I.V.  2003. Spin dynamics of terrestrial planets from
Earth-based RSDI,  Met. Planet. Sci.  38, Supplement, Abstract no. 5003.

Holsapple, K.A.  1993. The scaling of impact processes in planetary
sciences,  Ann. Rev. Earth Planet. Sci.  21, 333-373.

Kaula, W.M. 1966.  Theory of Satellite Geodesy; Applications of
Satellites to Geodesy, Blaisdell, Waltham, MA.

Knopoff, L. 1964. Q,  Rev. Geophys,  2, 625-660.

Lambeck, K. 1980.  The Earth's Variable Rotation: Geophysical
Causes and Consequences, Cambridge University Press, Cambridge.

Levison, H.F. Duncan, M.J. 1997. From the Kuiper Belt to Jupiter-family
comets: The spatial distribution of ecliptic comets,  Icarus 
127, 13-32.

Margot, J.L., Peale, S.J., Slade, M.A., Jurgens, R.F., Holin, I.V., 
2003. Mercury interior properties from measurements of librations. 
Mercury, 25th meeting of the IAU, Joint Discussion 2, 16 July, 2003, 
Sydney, Australia, meeting abstract. (http://adsabs.harvard.edu).

Peale, S.J. 1969. Generalized Cassini's laws,  Astron. J. 
74, 483-489.

Peale, S.J.  1974. Possible histories of the obliquity of Mercury,
Astron. J.  79, 722-744.

Peale, S.J.  1975. Dynamical consequences of meteorite impacts on the
Moon,  J. Geophys. Res.  80, 4939-4946. 

Peale, S.J.  1976a. Does Mercury have a molten core?  Nature 
262, 765-766. 

Peale, S.J.  1976b. Excitation and relaxation of the wobble,
precession and libration of the Moon,  J. Geophys Res.  81, 1813-1827.  

Peale, S.J.  1981. Measurement accuracies required for the
determination of a Mercurian liquid core,  Icarus  48,
143-145.

Peale, S.J.  1988. Rotational dynamics of Mercury and the state of its
core, In  Mercury, ed. F. Vilas, C.R. Chapman, M.S. Matthews,
U. of Arizona Press, Tucson, 461-493.

Peale, S.J. Phillips, R.J., Solomon, S.C., Smith, D.E., Zuber, M.T.,
 2002. A procedure for determining the nature of Mercury's core, 
Meteor. Planet. Sci.  37, 1269-1283.

Pike, R.J.  1988. Geomorphology of impact craters on Mercury, In 
Mercury, ed. F. Vilas, C.R. Chapman and M.S. Matthews, U. of Arizona
Press, Tucson. 165-273.

Rambaux, N., Bois, E. 2004. Theory of Mercury's spin-orbit motion
and analysis of its main librations,  Astron. Astrophys. 
413, 381-393.

Reese, C.C., Peterson, P.E. 2002. Thermal evolution of Mercury in
the conductive regime and implications for magnetism, 
Lun. Planet. Sci. XXXIII, Abstract 1998.

Siegfried, R.W., Solomon, S.C. 1974. Mercury: Internal structure
and thermal evolution,  Icarus  23, 192-205.

Smith, J.C., Born, G.H.  1976. Secular acceleration of Phobos and
the Q of Mars,  Icarus  27, 51-53.

Solomon, S.C., 20 colleagues  2001. The MESSENGER mission to Mercury:
Scientific objectives and implementation,  Planet. Space Sci.
 49, 1445-1465.

Spohn, T. and Schubert, G. 2003 Oceans in the icy Galilean
satellites of Jupiter? Icarus 161, 456-467.

Spohn, T.,  Sohl, F., Wieczerkowski, K. and Conzelmann, V. 2001 
The interior structure of Mercury: What we know, what we expect from
BepiColombo, Planet. Space Sci. 49, 1561-1570.  

Van Hoolst, T., Jacobs, C. 2003 Mercury's tides and interior
structure, J. Geophys. Res. 108, 7-(1-16).

Ward, W.M.  1975. Tidal friction and generalized Cassini's laws in the
solar system,  Astron. J.  80, 64-70.

Wijs, G.A. de, Kresse, G., Vo\~cadlo, L., Dobson, D., Alf\`e, D.,
Gillan, M.J., Price, G.D.  1998. The viscosity of liquid iron at the
physical conditions of the Earth's core,  Nature  392,
805-807.  

Zuber, M.T., Smith, D.E., 1997. Remote sensing of 
planetary librations from gravity and topography data: Mercury simulation.
Lunar Planet. Sci. 28, 1637-1638.
\newpage
\end{document}